\documentclass[fleqn,usenatbib]{mnras}

\usepackage{newtxtext,newtxmath}

\usepackage[T1]{fontenc}

\DeclareRobustCommand{\VAN}[3]{#2}
\let\VANthebibliography\thebibliography
\def\thebibliography{\DeclareRobustCommand{\VAN}[3]{##3}\VANthebibliography}

\usepackage{graphicx}	

\newcommand{\ph}{{\rm{ph}}}
\newcommand{\new}{{\rm{new}}}
\newcommand{\dd}{{\rm d}}
\newcommand{\lp}{\left(}
\newcommand{\rp}{\right)}
\newcommand{\beq}{\begin{equation}}
\newcommand{\eeq}{\end{equation}}
\newcommand{\Msun}{\ensuremath{\rm{M}_\odot}}
\newcommand{\Rsun}{\ensuremath{\rm{R}_\odot}}
\newcommand{\Lsun}{\ensuremath{\rm{L}_\odot}}
\newcommand{\Zsun}{\ensuremath{\rm{Z}_\odot}}
\newcommand{\mdot}{\ensuremath{\dot{M}}}

\title[A theory of mass transfer in binary stars]{A theory of mass transfer in binary stars}

\author[Cehula \& Pejcha]{
Jakub Cehula\thanks{E-mail: jakub.cehula@mff.cuni.cz} and
Ond\v{r}ej Pejcha
\\
Institute of Theoretical Physics, Faculty of Mathematics and Physics, Charles University, V Hole\v{s}ovi\v{c}k\'{a}ch 2, Prague, 180 00, Czech Republic
}

\date{Accepted XXX. Received YYY; in original form ZZZ}

\pubyear{2023}

\begin{document}
\label{firstpage}
\pagerange{\pageref{firstpage}--\pageref{lastpage}}
\maketitle

\begin{abstract}
Calculation of the mass transfer (MT) rate $\dot{M}_\text{d}$ of a Roche lobe overflowing star is a fundamental task in binary star evolution theory. Most of the existing MT prescriptions are based on a common set of assumptions that combine optically-thick and optically-thin regimes with different flow geometries. In this work, we develop a new model of MT based on the assumption that the Roche potential sets up a nozzle converging on the inner Lagrangian point and that the gas flows mostly along the axis connecting both stars. We derive a set of 1D hydrodynamic equations governing the gas flow with $\dot{M}_\text{d}$ determined as the eigenvalue of the system. The inner boundary condition directly relates our model to the structure of the donor obtained from 1D stellar evolution codes. We obtain algebraic solution for the polytropic equation of state (EOS), which gives $\dot{M}_\text{d}$ within a factor of 0.9 to 1.0 of existing optically-thick prescriptions and which reduces to the existing optically-thin prescription for isothermal gas. For a realistic EOS, we find that $\dot{M}_\text{d}$ differs by up to a factor of 4 from existing models. We illustrate the effects of our new MT model on $30\,M_\odot$ low-metallicity star undergoing intensive thermal time-scale MT and find that it is more likely to become unstable to L2 overflow and common-envelope evolution than for existing MT prescriptions. Our model provides a framework for including additional physics such as radiation or magnetic fields.
\end{abstract}




\section{Introduction}

Mass transfer (MT) between two stars within a binary system is a ubiquitous phenomenon in stellar astrophysics, which dominates the evolution of massive stars \citep{sana2012,demink2014,moe2017}. The idea of MT was first suggested by \citet{struve1941} to explain the variable spectrum of the eclipsing binary $\beta$ Lyr \citep{goodricke1785}. Since then, MT has been an accepted explanation for numerous astrophysical phenomena such as the Algol paradox in detached eclipsing binaries \citep[e.g.,][]{crawford1955,morton1960,paczynski1966,kippenhahn1967,pustylnik1998}, X-ray binaries \citep[e.g.,][]{hayakawa1964,zeldovich1966,pringle1972,shakura1973,boyle1984}, cataclysmic variable stars \citep[e.g.,][]{schatzman1958,kraft1964,smith2006}, or Type Ia supernovae \citep[e.g.,][]{wheeler1971,trimble1982,woosley1994sn}. 

In the era of gravitational-wave astronomy \citep[e.g.,][]{abbott2016observation,abbott2017gw170817,abbott2021}, MT is crucial for explaining populations of gravitational-wave mergers originating from isolated binary stars \citep[e.g.,][]{bhattacharya1991,belczynski2002,klencki2021,marchant2021,gallegos-garcia2021,gallegos-garcia2022,mandel2022,mandel2022b}. The two formation channels with MT are formation through stable MT or through common-envelope evolution. In stable MT, envelope of the donor is removed and orbital separation changes over a relatively long time interval. If the MT becomes unstable, the donor's envelope engulfs the accretor, the donor's core and the accretor spiral in inside a common envelope, and the surviving binary might be compact enough to merge within the Hubble time \citep[e.g.,][]{paczynski1976,ivanova2013,roepke2022}. The relative and absolute importance of the stable and unstable MT channels for various gravitational-wave progenitors is an actively debated topic. For example, \citet{gallegos-garcia2022} argued that binary neutron star formation may be dominated by common-envelope evolution, while \citet{pavlovskii2017}, \citet{vandenHeuvel2017}, \citet{gallegos-garcia2021}, \citet{klencki2021}, and \citet{marchant2021} reached an opposite conclusion for binary black holes.

More generally, the stability of MT and thus the onset of common-envelope evolution remains an unsolved problem.  One part of the issue is how does a star respond to mass removal and how important are thermal effects near the photosphere, where the thermal time-scale gets very short \citep[e.g.,][]{hjellming1987,podsiadlowski2002,ge2010,ge2015,ge2020,ge2023,woods2011,woods2012,passy2012,pavlovskii2015,temmink2022}. Another aspect of the problem is what happens to the gas after it flows near the inner Lagrangian point L1 into the domain of the companion. In the classical picture, if $| \mdot_\text{d} | \gtrsim 10^{-4}\, \Msun\,\rm{yr}^{-1}$, a large fraction of the infalling mass is lost to infinity through a fast, super-Eddington wind \citep[e.g.,][]{king1999}. Recently, \citet{lu2022} challenged this picture and suggested that the material is lost through the outer Lagrangian point L2, which is energetically more favorable. A similar outcome occurs when a star exceeds its Roche lobe radius so much that it overflows also the equipotentials passing through L2/L3. Mass lost through these points carries away large amount of angular momentum, which decreases the binary MT stability \citep[e.g.,][]{shu79,pejcha16a,pejcha16b,hubova19}. 

A central task in assessing stability of binary stars is the calculation of $\mdot_{\rm{d}}$ through L1/L2/L3 in 1D stellar evolution codes. Interestingly, the lineage of most of the currently used prescriptions can be traced to the Master thesis  of J\k{e}drzejec (1969) with the relevant results published by \citet{paczynski1972}. Following the works of \citet{plavec1973}, \citet{lubow1975}, \citet{savonije1978}, \citet{meyer1983}, \citet{ritter1988}, and others,  \citet{kolb1990} derived the currently widely used prescription for MT. Subsequent works mostly improved individual components of the model, but are based on the same assumptions (see Section~\ref{sec:review} for a review). Sometimes, only a simple exponential function of the Roche-lobe radius excess is used to calculate $\mdot_{\rm{d}}$ \citep[e.g.,][]{buning04,buning06}. A displeasing element of the existing theory is the separate treatment of the optically-thin (photosphere inside Roche lobe) and optically-thick (star overflows Roche lobe) regimes. When $\mdot_\text{d}$ is small, it does not significantly matter how is $\dot{M}_\text{d}$ calculated, because the star will simply overflow the Roche lobe slightly more or less to achieve the desired $\mdot_{\rm{d}}$. However, when $\mdot_{\rm{d}}$ is high and there is a complex interplay between dynamical, thermal, and orbital changes, the details of calculating $\mdot_{\rm{d}}$ matter. For example, if the star requires a high degree of Roche lobe overflow to get to the desired value of $\mdot_\text{d}$, its surface can overflow also L2/L3, which leads to a high rate of angular momentum loss from the binary. The procedure of calculating $\mdot_\text{d}$ is thus one of many systematic uncertainties in binary stellar evolution theory.

In this work, we develop a new model of MT through the L1 point. We abandon the standard assumption of utilizing the Bernoulli's principle to evolve gas flowing from donor's surface toward L1 \citep{lubow1975,ritter1988,kolb1990}. Instead, we assume that the Roche potential sets up a converging-diverging (de Laval-like) nozzle around L1. Although the Roche potential sets up the geometry, fluid elements are allowed to cross equipotentials. We describe the gas flowing through the nozzle using time-steady Euler equations averaged over the plane perpendicular to the gas motion. By applying a set of assumptions,  we obtain 1D two-point boundary value problem starting at a point below the donor's surface and ending at L1. The MT rate is the eigenvalue of the problem, similarly to stellar wind calculations. At this stage of our work we are only able to account for adiabatic gas flow, but the advantage of our model is that eventually we will be able to implement additional physics such as radiation transport or magnetic fields. Our model can be included in 1D stellar evolution codes without a significant increase in computational time.

This paper is organised as follows: in Section~\ref{sec:review}, we review existing models of MT. In Section~\ref{sec:hydro}, we present the derivation of our new model. In Section~\ref{sec:solution}, we describe the methods of solution for ideal gas and a more realistic equation of state (EOS). In Section~\ref{sec:results}, we apply our new model to a Sun-like donor on main sequence or red giant branch and to a low-metallicity massive donor losing mass on thermal time-scale. In Section~\ref{sec:comparison}, we compare our new model with the existing models. We also illustrate the effects of our new MT model on binary star evolution. In Section~\ref{sec:discussion}, we discuss and summarize our results.

\section{Review of existing mass-transfer models}\label{sec:review}

In this section, we sum up existing  models of MT. In Section~\ref{sec:framework}, we introduce a general framework for all MT models. In Section~\ref{sec:opt_thin}, we summarize models of optically-thin MT when the donor's photosphere lies within the Roche lobe. In Section~\ref{sec:opt_thick}, we review models of optically-thick MT for donor that overflows the Roche lobe.

\subsection{Framework}\label{sec:framework}
MT is commonly studied in Roche geometry (Fig.~\ref{fig:bmt_scheme_KR}) described by the Roche potential $\phi_\text{R}$. In this framework, we denote the masses of the donor and accretor by $M_{\rm{d}}$ and $M_{\rm a}$. The total mass of the binary is $M = M_{\rm d} + M_{\rm a}$ and the mass ratio  $q = M_{\rm d}/M_{\rm a}$. The binary angular velocity is $\omega^2 = GM/a^3$, where $G$ is the gravitational constant and $a$ is the binary separation. We assume that the orbital plane is the $xy$ plane and that the centres of the two stars are on the $x$ axis. The donor has $x<0$, the accretor has $x>0$ and L1 is in the centre of the coordinate system, $x_1 = 0$. The lowest order approximation of the Roche potential around L1 is
\begin{equation}
    \phi_{\rm R} (x,y,z) - \phi_1 \approx -\frac{A}{2} x^2 + \frac{B}{2} y^2 + \frac{C}{2} z^2,
    \label{eq:phi_app}
\end{equation}
where $A$, $B$, and $C$ are positive constants and $\phi_1$ is the Roche potential at L1. Following \citet{eggleton1983approximations}, we approximate the donor's Roche-lobe radius as $R_{\rm{L}} = f (q) a$, where
\begin{equation}
    f (q) = \frac{0.49 q^{2/3}}{0.6q^{2/3} + \ln ( 1 + q^{1/3})}.
\end{equation}
The relative radius excess of the donor $\delta R_\text{d}$ is
\begin{equation}
    \delta R_{\rm d} \equiv \frac{\Delta R_{\rm d}}{R_{\rm{L}}} = \frac{R_{\rm d} - R_{\rm{L}}}{R_{\rm{L}}},
\end{equation}
where $\Delta R_{\rm{d}} \equiv R_{\rm d} - R_{\rm{L}}$ is the radius excess and $R_{\rm d}$ is the radius of the donor's photosphere. Relative radius excess $\delta R_\text{d}$ is the crucial quantity on which the MT rate $\dot{M}_{\rm d}$ of the donor through L1 depends. Traditionally, if the donor is Roche-lobe underfilling, $\delta R_{\rm d} <0$, MT is treated as optically-thin and isothermal \citep{lubow1975, ritter1988, jackson2017}. If the donor is Roche-lobe overflowing, $\delta R_{\rm d} >0$, MT is treated as optically thick and adiabatic \citep{kolb1990, pavlovskii2015, marchant2021}.

\subsection{Optically-thin mass transfer}\label{sec:opt_thin}
The treatment of stationary optically-thin MT was developed by \citet{ritter1988} using the previous results of \citet{lubow1975}. The prescription is based on the fact that even if $\delta R_{\rm d} < 0$, the density profile of donor's atmosphere extends above its photosphere to L1. The gas above the photosphere can be considered optically-thin to radiation coming from the donor, but simultaneously the gas achieves equilibrium with the surrounding radiation field. Before leaving through L1, the gas stays in close proximity of the donor's surface, where the radiation field is roughly constant. As a result, the MT flow can be treated as isothermal at temperature $T \approx T_{\rm{eff}}$, where $T_{\rm{eff}}$ is the donor's effective temperature. After the gas passes L1, it is accreted supersonically onto the accretor. \citet{lubow1975} derived that the gas reaches the isothermal sound speed $c_T$ close to L1. \citet{ritter1988} expressed the MT rate as
\begin{equation}
    - \dot{M}_{\rm d} = \rho_1 c_T Q_\rho,
    \label{eq:Mdot_R}
\end{equation}
where $\rho_1$ is the gas density at L1 and $Q_\rho$ is the effective cross-section corresponding to the gas density profile at L1. The quantity $\dot{M}_{\rm{d}}$ is negative because the donor is losing mass.

In order to evaluate the density at L1, \citet{ritter1988} used a form of Bernoulli's equation originally derived by \citet{lubow1975},
\begin{equation}
    \frac{1}{2} v^2 + \int \frac{dP}{\rho} + \phi_{\rm R} = \text{constant along a streamline},
    \label{eq:Bernoulli}
\end{equation}
where $v$ is the gas velocity and $P$ is the pressure. To evaluate equation~(\ref{eq:Bernoulli}), \citet{ritter1988} used the assumption of ideal gas,
\beq
    P = \frac{k}{\overline{m}} T \rho = c_T^2 \rho, \quad \epsilon = \frac{1}{\Gamma-1} \frac{k}{\overline{m}} T = \frac{1}{\Gamma-1} c_T^2,
    \label{eq:id}
\eeq
where $k$ is the Boltzmann constant, $\overline{m}$ is the mean mass of a gas particle, $\epsilon$ is the internal energy per unit mass, and $\Gamma = c_P / c_V$, where $c_P$ and $c_V$ are dimensionless heat capacities at constant pressure and volume. By assuming that the gas starts with negligible velocity at donor's photosphere, $v_{\rm{ph}}^2 \ll c_T^2$, equations~(\ref{eq:Bernoulli}) and (\ref{eq:id}) provide
\begin{equation}
    \rho_1 = \frac{1}{\sqrt{\rm e}} \rho_{\rm{ph}} \exp \lp -\frac{\phi_1-\phi_{\rm{ph}}}{c_T^2}\rp,
    \label{eq:rho_1}
\end{equation}
where $\rho_\ph$ and $\phi_\ph$ are the density and the Roche potential of donor's photosphere. 

To evaluate the potential difference $\phi_1 - \phi_\ph$, we need to know which equipotential $\phi_V$ is reached by a star with radius $r_V$. Here, volume-equivalent radius $r_V$ is defined using a ball with a volume $V$ enclosed by the equipotential surface on the donor's side of the L1 plane, $V (\phi_V) = 4 \pi r_V^3/ 3$. The L1 plane is the $yz$ plane going through L1. \citet{ritter1988} used first-order expansion of $\phi_V(r_V)$, but here we use higher-order expansion derived by \citet{jackson2017},
\begin{equation}
    \begin{aligned}
        & \phi_V (r_V) = -G \frac{M_{\rm a}}{a} \left[ 1+ \frac{1}{2\lp 1+ q\rp}\right] \\
        & \quad -G \frac{M_{\rm d}}{r_V} \left[ 1+ \frac{1}{3} \lp 1+ \frac{1}{q}\rp \lp \frac{r_v}{a}\rp^3 + \frac{4}{45} \lp 1 + \frac{5}{q} + \frac{13}{q^2}\rp \lp \frac{r_v}{a}\rp^6 \right].
    \end{aligned}
    \label{eq:phi_vol}
\end{equation}
This expression converges inside donor's Roche lobe, $r_V \leq R_{\rm{L}}$, for $10^{-3} \le q \le 10^2$ to within 2 per cent. With these definitions,  the potential difference can be evaluated as $\phi_1 - \phi_\ph = \phi_V( R_{\rm{L}}) - \phi_V (R_{\rm d})$.

To evaluate $Q_\rho$, \citet{ritter1988} followed \citet{meyer1983} who calculated the hydrostatic isothermal drop-off of the density in the L1-plane using the potential approximation in equation~(\ref{eq:phi_app}). In our notation, we obtain
\begin{equation}
    Q_\rho = \frac{2 \pi}{\sqrt{BC}} c_T^2.
    \label{eq:Q_rho}
\end{equation}
\citet{ritter1988} provided approximate expressions for $A$, $B$, and $C$ valid for $0.1 \le q \le 2$. Here, we use improved relations of \citet{jackson2017},
\begin{equation}
    \begin{aligned}
        &A = (1 + 2 \mathcal{A}) \omega^2, \quad B = (\mathcal{A} - 1) \omega^2, \quad C = \mathcal{A} \omega^2, \\
        & \quad \mathcal{A} = 4 + \frac{4.16}{-0.96 + q^{1/3} + q^{-1/3} },
    \end{aligned}
    \label{eq:A}
\end{equation}
where the expression for $\mathcal{A}(q)$ approximates the true value to better than 0.3 per cent.

To summarize, the optically-thin MT rate with the improvements of \citet{jackson2017}, $\dot{M}_\text{J}$, is
\begin{equation}
    - \dot{M}_{\rm d} \equiv \dot{M}_{\rm J} = \dot{M}_{\rm{J,0}} \exp \lp - \frac{\phi_1-\phi_\ph}{c_T^2}\rp, \quad \dot{M}_{\rm{J,0}} = \frac{2\pi}{\sqrt{\rm e}} \frac{1}{\sqrt{BC}} c_T^3 \rho_\ph.
    \label{eq:Mdot_J}
\end{equation}
In this regime, the gas is assumed to flow across the equipotentials (equation~\ref{eq:Bernoulli}) and the procedure used in this calculation is similar to the one employed in estimating the mass-loss rate of a spherical isothermal wind \citep[Sec.~3 and fig.~3.2]{lamers1999}. The isothermal assumption could be violated if the gas is moving fast enough so that thermal equilibrium with radiation is not achieved or by chromospheric or coronal heating, which increase $c_T$, the density scale height, and correspondingly the MT rate.

\subsection{Optically-thick mass transfer}\label{sec:opt_thick}

\begin{figure}
	\includegraphics[width=\columnwidth]{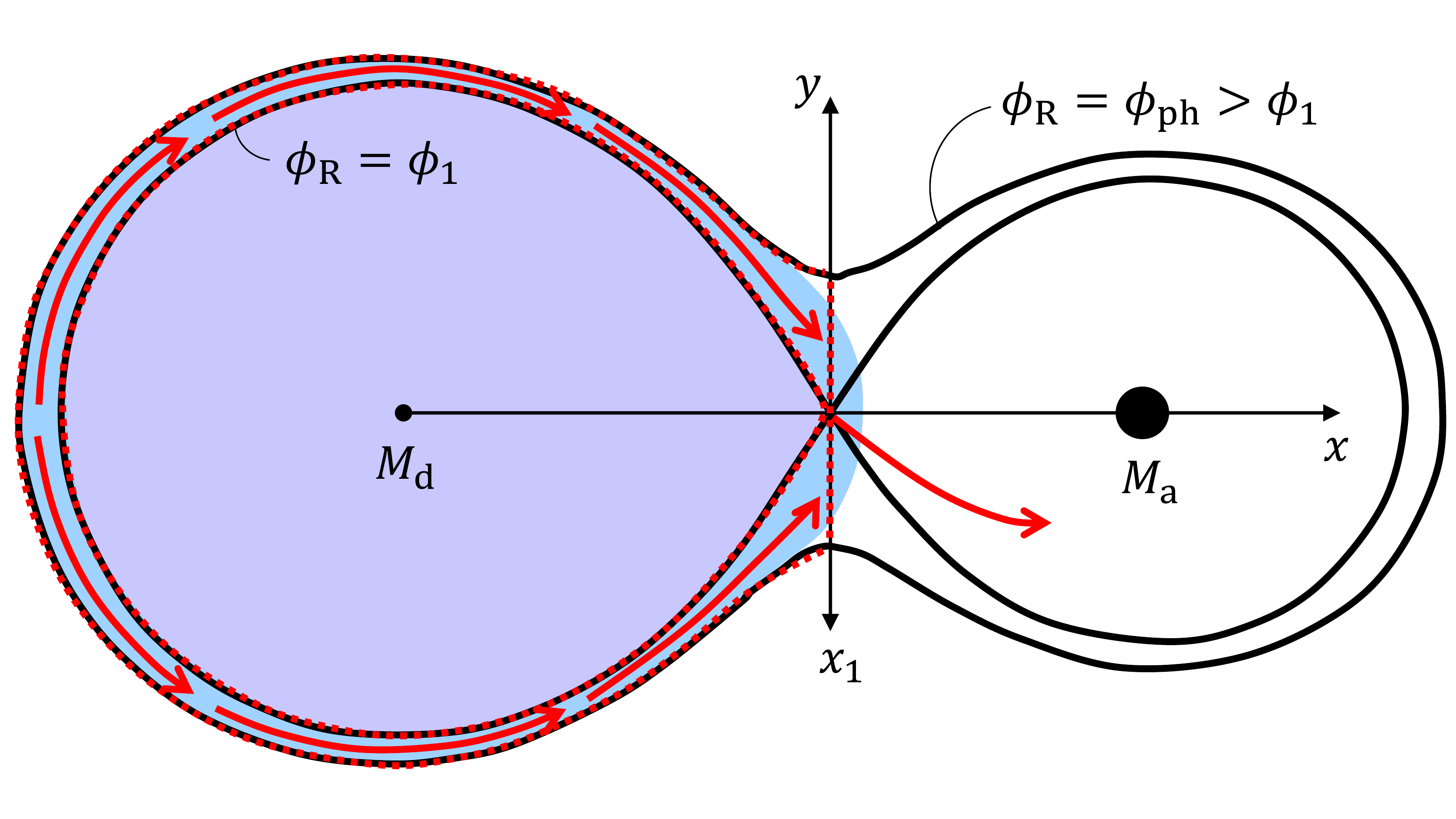}
    \caption{Diagram of gas flow in the optically-thick model of \citet{kolb1990}. The equipotentials of Roche geometry are indicated by thick black lines. The gas in hydrostatic equilibrium is shown in light purple, while the gas transferred to the accretor is marked in light blue. The modelled region is indicated by the red dotted line and the red arrows indicate the direction of gas flow.}
    \label{fig:bmt_scheme_KR}
\end{figure}

\citet{kolb1990} derived the currently standard way of modeling stationary MT in the optically-thick regime. In Fig.~\ref{fig:bmt_scheme_KR}, we schematically show the assumed flow of gas. In this model, the gas below the photosphere is assumed to be optically-thick, and, more importantly, it is assumed that there is no energy transport within the gas and between the gas and the surrounding radiation field. As a result, the MT flow is adiabatic. Gas with potential $\phi < \phi_1$ remains in hydrostatic equilibrium whereas the gas with higher potential, $\phi_1 < \phi < \phi_\ph$, flows toward L1 along streamlines that lie nearly on equipotential surfaces, as was argued by \citet{lubow1975}. Similarly to the optically-thin case, the gas reaches adiabatic sound speed $c_s$ in the close vicinity of L1. The MT rate is given by integration over all streamlines,
\begin{equation}
    - \dot{M}_{\rm{d,thick}} = \int_{\rm{L1-plane}} \rho_{\rm L} c_s \dd Q,
    \label{eq:Mdot_thick}
\end{equation}
where $\rho_{\rm L}$ is the density in the L1 plane and $\dd Q$ is the area of a streamline. Both $\rho_{\rm L}$ and $c_s$ vary across the L1 plane.

Since the gas flow is assumed to be adiabatic, \citet{kolb1990} used polytropic EOS along a streamline,
\begin{equation}
    P = K \rho^\Gamma, \quad c_s^2 = \Gamma c_T^2 = \Gamma \frac{P}{\rho},
    \label{eq:polytrope}
\end{equation}
where $K$ is the polytropic constant and $\Gamma$ is the polytropic exponent. To evaluate $\rho_{\rm L}$, the starting point of the streamline is positioned somewhere far away from L1 where hydrostatic equilibrium can be assumed and the starting density is given by hydrostatic stellar structure. The endpoint of the streamline is in the L1 plane where $v=c_s$. \citet{kolb1990} assumed that $\phi_{\rm R}$ is constant along the streamline and that the gas starts with a negligible velocity. Combining equation~(\ref{eq:Bernoulli}) with the polytropic assumption in equation~(\ref{eq:polytrope}) gives
\begin{equation}
    \rho_{\rm L} (\phi_{\rm R}) = \lp \frac{2}{\Gamma+1} \rp^{\frac{1}{\Gamma-1}} \bar{\rho} ( \bar{\phi}  ),
\end{equation}
where $\bar{\rho}$ denotes the hydrostatic density given by 1D stellar evolution code evaluated at the hydrostatic potential $\bar{\phi} = \phi_\text{R}$. 

Finally, \citet{kolb1990} combined equation~(\ref{eq:Mdot_thick}) with the expression for saturated optically-thin MT rate (equation~\ref{eq:Mdot_J}) to obtain the final expression for optically-thick MT rate, $\dot{M}_\text{KR}$,
\begin{equation}
    -\dot{M}_{\rm d} \equiv \dot{M}_{\rm{KR}} = \dot{M}_{\rm{J,0}} +  \left. \frac{\dd Q}{\dd \phi} \right\vert_{\rm{L1}} \int_{\phi_1}^{\phi_\ph} F_3 \lp\Gamma \rp \lp \frac{k \bar{T}}{\overline{m}}\rp^{\frac{1}{2}} \bar{\rho} \dd \bar{\phi},
    \label{eq:Mdot_KR}
\end{equation}
where $\bar{T}$ is the hydrostatic temperature and
\begin{equation}
    F_3 \lp \Gamma\rp = \Gamma^{\frac{1}{2}} \lp \frac{2}{\Gamma+1}\rp^{\frac{\Gamma+1}{2\lp\Gamma-1\rp}}, \quad \left. \frac{\dd Q}{\dd \phi} \right\vert_{\rm{L1}} = \frac{2\pi}{\sqrt{BC}}.
    \label{eq:F3}
\end{equation}
Here, the potential increment in the 1D hydrostatic model is $\dd \bar{\phi} = - \dd \bar{P} / \bar{\rho}$, where $\bar{P}$ is the hydrostatic pressure. 

Significant improvements of equation~(\ref{eq:Mdot_KR}) were made by \citet{pavlovskii2015} and \citet{marchant2021}. First, $\Gamma$ is not necessarily a constant along a streamline, because of the changing ionization. The effect of varying $\Gamma$ on $\dot{M}_{\rm{d,thick}}$ is $\lesssim $4 per cent \citep{pavlovskii2015}. Second, the ideal gas assumption can be violated for layers with significant radiation pressure. Third, it is possible to use higher-order expansion of the Roche potential in the L1 plane than what is provided by equation~(\ref{eq:phi_app}). In such situation, $\dd Q / \dd \phi \neq (\dd Q / \dd \phi)_{\rm{L1}} = \rm{const.}$ and this term cannot be taken out of the integral in equation~(\ref{eq:Mdot_KR}). This improvement is useful when the donor significantly overflows L1 such that it may lose mass via L2/L3. To facilitate easier comparison of our new mass transfer model to the model of \citet{marchant2021} in Sections~\ref{sec:30-Msun_comp}, \ref{sec:comp_evol}, we denote their MT rate by $\dot{M}_{\rm{M}}$.

\section{New mass-transfer model}\label{sec:hydro}

In this Section, we develop a MT model that works in both isothermal and adiabatic regimes at the same time. Before deriving model equations (Sec.~\ref{sec:model_equations}) and comparison with previous efforts along similar lines (Sec.~\ref{sec:similarities}), we shortly explain here the main idea. 

It is difficult to imagine that the MT flow streamlines would suddenly change from crossing the equipotentials to being aligned with equipotentials once the photosphere moves outside of the Roche lobe (Fig.~\ref{fig:bmt_scheme_KR}). Instead, we think that it is more natural to interpret the MT flow as a special case of a non-spherical stellar wind in a 3D potential. This analogy leads us to postulate that MT streamlines can cross equipotentials in both regimes. Additionally, the MT rate should be viewed as an eigenvalue of a hydrodynamical configuration composed of the entire sonically connected stellar body both below and above the Roche lobe. Practically, this implies recalculating the hydrodynamic structure of the star in the vicinity of the L1 point.

\begin{figure}
	\includegraphics[width=\columnwidth]{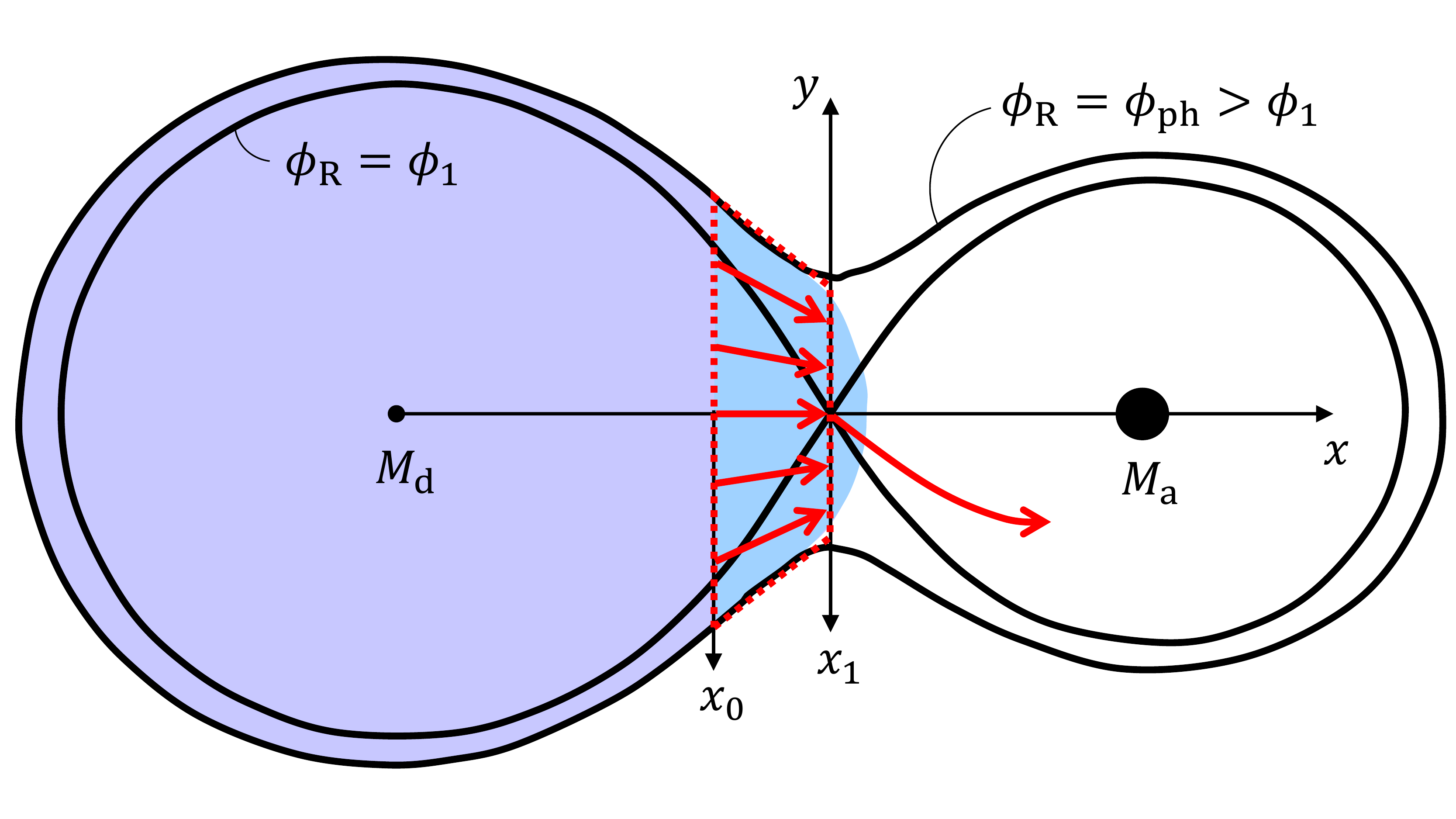}
    \caption{Diagram of gas flow in our model of MT. The meaning of the symbols is the same as in Fig.~\ref{fig:bmt_scheme_KR}.}
    \label{fig:bmt_scheme}
\end{figure}

In Fig.~\ref{fig:bmt_scheme}, we show a schematic diagram of the MT flow in our model. We illustrate our ideas on a Roche-lobe overflowing donor, where the potential in donor's photosphere is greater than the potential at L1, $\phi_{\rm{ph}} > \phi_1$. We assume that the Roche potential sets up a de Laval-like nozzle around L1. The gas flows mainly along the $x$ axis and we recalculate donor's hydrodynamic structure from some point below the donor's surface $x_0$ up to the L1 point at $x_1 = 0$. The $x_0$ point connects our calculation to the underlying 1D hydrostatic stellar model, however, the exact position of $x_0$ will be determined later. The gas reaches critical speed at the narrowest point of the nozzle at L1. From L1 the gas falls freely onto the accretor.

\subsection{Model equations}
\label{sec:model_equations}
We start with 3D Euler hydrodynamic equations including the Roche potential, but neglecting Coriolis force, viscosity, magnetic fields, and radiation. The time evolution of mass, linear momentum, and energy are
\begin{subequations}
	\label{eq:hydro}
	\begin{align}
	\frac{\partial \rho}{\partial t} + \boldsymbol{\nabla} \boldsymbol{\cdot} (\rho \boldsymbol{v}) &= 0 , \\
	\frac{\partial (\rho \boldsymbol{v} )}{\partial t} + \boldsymbol{\nabla} \boldsymbol{\cdot}(\rho \boldsymbol{v} \otimes \boldsymbol{v} + P \boldsymbol{\mathsf{I}} ) &= -\rho \boldsymbol{\nabla} \phi_{\rm R}, \\
	\frac{\partial ( \rho \epsilon_{\rm{tot}} )}{\partial t} + \boldsymbol{\nabla} \boldsymbol{\cdot} \left[ (\rho \epsilon_{\rm{tot}} + P ) \boldsymbol{v}\right] &= 0,
	\end{align}
\end{subequations}
where $\boldsymbol{v}$ is the 3D velocity vector, $\boldsymbol{\mathsf{I}}$ is the unit matrix, and $\epsilon_{\rm{tot}}$ is the total energy per unit mass, 
\begin{equation}
\epsilon_{\rm{tot}} = \epsilon + \frac{1}{2} \boldsymbol{v}^2 + \phi_{\rm R},
\label{eq:eps_tot}
\end{equation}
which consists of internal, kinetic, and potential components.

To turn the 3D Euler equations in a 1D eigenvalue problem, we apply these assumptions:
\begin{enumerate}
    \item The gas flow is stationary, $\partial / \partial t \longrightarrow 0$.
    \item The $x$ component of the velocity does not depend on the position in the $yz$ plane, $v_x( x, y, z) = v_x(x) $.
    \item The flow proceeds mainly along the $x$ axis, $v_y^2 \ll v_x^2$, $v_z^2 \ll v_x^2$, which implies $\boldsymbol{v}^2 \approx v_x^2$. This assumption also allows us to use hydrostatic equilibrium in the $yz$ plane, 
    \begin{equation}
        \left.\frac{1}{\rho} \nabla P\right\vert_{x=\rm{const}} = -\left.\nabla \phi_{\rm R} \right\vert_{x=\rm{const}}.
    \end{equation}
    \item The Roche potential in the proximity of L1 can be split in two parts: one depends solely on $x$, $\phi_{\rm R}^x$, and the other one on $y$ and $z$, $\phi_{\rm R}^{yz}$. We use the lowest-order approximation for $\phi_{\rm R}^{yz}$, 
    \begin{equation}
    \phi_{\rm R}(x, y, z) = \phi_{\rm R}^x(x) + \phi_{\rm R}^{yz}(y, z) = \phi_{\rm R}^x(x) + \frac{B}{2} y^2 + \frac{C}{2} z^2,
    \end{equation}
    which directly implies  that $\dd \phi_{\rm R}/\dd x = \dd \phi_{\rm R}^x/\dd x$.
    \item The gas is polytropic in the $yz$ plane:
    \begin{equation}
        \left.P \right\vert_{x=\rm{const}} = \left.K\rho^\Gamma \right\vert_{x=\rm{const}},
        \nonumber
    \end{equation}
    where $K=K(x)$ and $\Gamma = \Gamma(x)$.
\end{enumerate}
This set of assumptions is not unique or necessarily better than what was used in previous models, however, they provide a straightforward way to formulate a 1D problem.

In order to effectively apply our assumptions, we also need to define the effective cross-section of the gas in the $yz$ plane, $Q(x)$. Our assumption (v) guarantees that the density away from the $x$ axis eventually reaches zero so that we can define the cross-section by the expression $\rho(\partial Q) = 0$, where $\partial Q$ is the boundary of the cross-section. We define the effective cross-sections $Q_\rho$ and $Q_P$ corresponding to the density and pressure profiles in the $yz$ plane as
\begin{subequations}
    \label{eq:Q_def}
    \begin{align}
    \rho (x) Q_\rho &= \rho(x,0,0) Q_\rho (x) \equiv \int_{Q(x)} \rho (x,y,z) \dd Q, \label{eq:Qrho} \\
    P(x) Q_P &= P(x,0,0) Q_P(x) \equiv \int_{Q(x)} P (x,y,z) \dd Q. \label{eq:QP}
    \end{align}
\end{subequations}
In principle, to get finite $Q_\rho$ and $Q_P$ we only need a density profile that decreases sufficiently fast away from the $x$ axis. The polytropic assumption (v) provides a convenient way to estimate these cross-sections, but it is not uniquely needed.

Next, we integrate the 3D Euler equations~(\ref{eq:hydro}) over the $yz$ plane. The derivation is detailed in Appendix~\ref{sec:hydro_gen} and the final set of 1D hydrodynamic equations is
\begin{subequations}
\label{eq:hydro_gen}
	\begin{align}
	\frac{1}{v}\frac{\dd v}{\dd x} + \frac{1}{\rho Q_\rho}\frac{\dd}{\dd x} (\rho Q_\rho) &= 0, \label{eq:hydro_gen_mass} \\
	v\frac{\dd v}{\dd x} + \frac{1}{\rho Q_\rho} \frac{\dd}{\dd x} (P Q_P) &= -\frac{\dd\phi_{\rm R}}{\dd x}, \label{eq:hydro_gen_mom} \\
	\frac{\dd}{\dd x} \left( \epsilon \frac{Q_P}{Q_\rho}\right) - \frac{PQ_P}{(\rho Q_\rho)^2} \frac{\dd}{\dd x}(\rho Q_\rho) &= - \frac{\dd}{\dd x} \left( c_T^2 \frac{Q_P}{Q_\rho}\right), \label{eq:hydro_gen_en}
	\end{align}
\end{subequations}
where $v$ now stands for $v_x$ and all the variables are evaluated on the $x$ axis and are functions of $x$, e.g., $\epsilon = \epsilon(x) = \epsilon(x,0,0)$. For the effective density cross-section $Q_\rho$, we recover equation~(\ref{eq:Q_rho}). The ratio of the two cross-sections is 
\begin{equation}
    \frac{Q_P}{Q_\rho} = \frac{\Gamma}{2\Gamma-1}.
    \label{eq:Q_frac}
\end{equation} 
The individual terms in equations~(\ref{eq:hydro_gen}) have meaning similar to the terms in the original equations~(\ref{eq:hydro}). One exception is the new term on the right-hand side of equation~(\ref{eq:hydro_gen_en}), which captures the changes in potential energy from compression or expansion in the $yz$ plane.  
Finally, equation~(\ref{eq:hydro_gen_mass}) expresses conservation of mass flux along the $x$ axis so that we can express the MT rate of donor through L1, $\dot{M}_\text{new}$,\begin{equation}
    - \dot{M}_{\rm d} \equiv \dot{M}_{\rm{new}} = v \rho Q_\rho = \frac{2\pi}{\sqrt{BC}} c_T^2 v \rho.
    \label{eq:Mdot}
\end{equation}

In order to close the system of hydrodynamic equations~(\ref{eq:hydro_gen}) we need an EOS, which provides $c_T$, $P$, $\epsilon$, $\Gamma$ as a function of $\rho$ and $T$. Thus, the hydrodynamic equations~(\ref{eq:hydro_gen}) become a two-point boundary value problem for three unknowns, $v$,  $\rho$, and $T$. We have two boundary conditions at $x_0$, $\rho(x_0) = \rho_0$ and $T(x_0) = T_0$, which are fixed to values from a 1D hydrostatic model. The third boundary condition is applied at $x_1$, $v(x_1) = v_1(\rho(x_1), T(x_1))$, which is the critical point. In order to evaluate the boundary condition at $x_1$, we derive the matrix form of hydrodynamic equations~(\ref{eq:hydro_gen}),
\begin{equation}
    \boldsymbol{\mathsf{M}}
    \begin{pmatrix}
        \frac{1}{v}\frac{\dd v}{\dd x} & \frac{1}{\rho}\frac{\dd \rho}{\dd x} & \frac{1}{T}\frac{\dd T}{\dd x}
    \end{pmatrix}^\top
    =
    \begin{pmatrix}
        0 & -\frac{\dd\phi_{\rm R}}{\dd x} & 0
    \end{pmatrix}^\top,
    \label{eq:hydro_gen_mat}
\end{equation}
where the exact form of the matrix $\boldsymbol{\mathsf{M}}$ is shown in Appendix~\ref{sec:hydro_gen_mat}. At the critical point $x_1$, it holds that $\dd\phi_{\rm R}/\dd x = 0$, therefore the boundary condition is $\det \boldsymbol{\mathsf{M}} (x_1) = 0$, and the vector on the left-hand side of equation~(\ref{eq:hydro_gen_mat}) belongs to the kernel of $\boldsymbol{\mathsf{M}}(x_1)$.

\subsection{Similarities to previous models}
\label{sec:similarities}
An approach similar to our model was already considered by \citet{nariai1967} who was motivated to explain periodically appearing blue-shifted H$\alpha$ absorption line in the spectrum of $\upsilon$~Sgr. \citet{nariai1967} suggested that the Roche potential sets up a de Laval nozzle around L1 and that the coronal wind of one of the stars passes through the nozzle, where it reaches supersonic velocities. This supersonic wind was supposed to explain periodic coverage of the binary and thus periodically appearing blue-shifted H$\alpha$ absorption line\footnote{The possible explanation of the peculiar spectrum of $\upsilon$~Sgr suggested by \citet{nariai1967} was ruled out by \citet{koubsky2006} based on radial velocity measurements.}. The most important difference between \citet{nariai1967} and our approach is that we include the Roche potential in our model and \citet{nariai1967} assumes polytropic EOS.

\citet{lubow1975} had two objections to \citet{nariai1967}. First, corona-like temperatures are required to explain the displacement of the H$\alpha$ absorption line. Second, de Laval nozzle enforces converging-diverging streamlines but converging-diverging equipotentials around L1 do not. In our model, we do not rely on corona-like temperatures to explain the MT. To the second objection, gas flow in our MT model is readily allowed to cross equipotentials both naturally along the $x$ axis, but also in the $yz$ plane, where we assume hydrostatic density structure with scale height depending on the temperature on the $x$ axis.

\section{Method of solution}\label{sec:solution}

The method of solving the set of 1D hydrodynamic equations~(\ref{eq:hydro_gen}) depends on the EOS we use. In Section~\ref{sec:eos_isothermal}, we formulate our model for isothermal gas, which leads to an algebraic solution. In Section~\ref{sec:eos_ideal}, we show the same but for ideal gas. Finally, in Section~\ref{sec:eos_real}, we describe numerical procedure for solving our equations for an arbitrary EOS.

\subsection{Isothermal gas}
\label{sec:eos_isothermal}
We start with a simple isothermal gas with $T = T_0 = {\rm{const}}$, $c_T = {\rm{const}}$, which is the limiting case of ideal gas if $\Gamma \rightarrow 1$ (equation \ref{eq:hydro_id}). The hydrodynamic equations are
\begin{subequations}
	\label{eq:hydro_iso}
	\begin{align}
	\frac{\dd \ln v}{\dd x} + \frac{\dd \ln \rho}{\dd x} &= 0, \label{eq:hydro_iso_mass} \\
	v^2\frac{\dd \ln v}{\dd x} + c_T^2 \frac{\dd \ln \rho}{\dd x} &= -\frac{\dd\phi_{\rm R}}{\dd x}. \label{eq:hydro_iso_mom} \\
	T &= T_0 = {\rm{const}}, \label{eq:hydro_iso_en}
	\end{align}
\end{subequations}
We now effectively have a set of two equations for two unknowns, $v$ and $\rho$. The boundary conditions are: $\rho (x_0) = \rho_0$ and $v (x_1) = c_T$. We solve these equations in a form of a set of algebraic relations in Section~\ref{sec:res_iso}.

\subsection{Ideal gas}
\label{sec:eos_ideal}
For ideal gas described by equation~(\ref{eq:id}) we have
\begin{subequations}
	\label{eq:hydro_id}
	\begin{align}
	\frac{\dd \ln v}{\dd x} + \frac{\dd \ln \rho}{\dd x} + \frac{\dd \ln T}{\dd x} &= 0, \label{eq:hydro_id_mass} \\
	v^2\frac{\dd \ln v}{\dd x} + \Gamma c_T^2 \frac{\dd \ln \rho}{\dd x} &= -\frac{\dd\phi_{\rm R}}{\dd x}, \label{eq:hydro_id_mom} \\
	\frac{\dd \ln \rho}{\dd x} - \frac{1}{\Gamma-1} \frac{\dd \ln T}{\dd x} &= 0. \label{eq:hydro_id_en}
	\end{align}
\end{subequations}
The boundary conditions are: $\rho (x_0) = \rho_0$, $T (x_0) = T_0$, and $v^2 (x_1) = c_T^2 (x_1) = k T ( x_1) / \overline{m}$. This set of differential equations can be further reduced to a set of algebraic equations, as we show in detail in Appendix~\ref{sec:hydro_id}. We discuss the solution in Section~\ref{sec:res_id}.

\subsection{Realistic equation of state}
\label{sec:eos_real}
For a general EOS, we solve the set of hydrodynamic equations~(\ref{eq:hydro_gen}) with relaxation code based on \citet[Chapter 18]{press2007}. Numerical solution is facilitated by realizing that there is no need to use the physical $x$ coordinate as an independent variable, which requires a specific approximation of the Roche potential on the $x$ axis, $\phi_{\rm R}^x$. Consequently, we use $\phi_{\rm R}^x$ as independent variable which ranges from $\phi_0$ to $\phi_1$, where $\phi_0 \equiv \phi_{\rm R}^x (x_0)$ and $\phi_1 \equiv \phi_{\rm R}^x ( x_1)$. To improve convergence, we use logarithmic scaling of the quantities, which makes our dependent variables $\log v$, $\log \rho$, and $\log T$. We obtain best convergence with a quadratic grid of the independent variable defined as
\begin{equation}
    \phi [j] = - \frac{\phi_1-\phi_0}{N_\text{g}^2} j^2 + 2 \frac{\phi_1-\phi_0}{N_\text{g}} j + \phi_0,
\end{equation}
where $\phi[j]$ is the $j$-th grid point and $j = 0,1,2,...,N_\text{g}$. We usually use $N_\text{g} = 200$. In order to compute numerical derivatives of the expressions in equations~(\ref{eq:hydro_gen}), we use symmetric derivative with differentiation of the variables by 0.5 per cent. The boundary conditions for our calculation are $\rho (\phi_0) = \rho_0$, $T (\phi_0) = T_0$, and $\det \boldsymbol{\mathsf{M}} (\phi_1) = 0$. 

Now we address the question how to obtain $\rho_0$ and $T_0$ as functions of the difference $\phi_1-\phi_0$. Typically, the donor is modelled using some 1D stellar evolution code, which provides profiles of $\rho$, $T$, $P$, and similar quantities at each time-step as functions of the radius~$R$. This leaves us with the problem of finding the mapping $\phi^x_{\rm{R}} (R)$, specifically, the value $\phi_0 (R_0)$ for some radius $R_0$ below the donor's photosphere (see Sec.~\ref{sec:potential_ambiguity} for discussion). We can then read the values $\rho_0$ and $T_0$ simply as $\rho_0 = \bar{\rho} (R_0)$ and $T_0 = \bar{T} (R_0)$. In the case of Roche-lobe overflowing donor, $\delta R_{\rm d}>0$, the radius $R_{\rm{L}}$ is within the donor. Thus, the value of the potential $\phi_0$ corresponds to the radius $R_0$ if and only if we define $R_0$ in the following way
\begin{equation}
    \phi_1 - \phi_0 \equiv \int_{R_0}^{R_{\rm{L}}} \dd \bar{\phi} = - \int_{R_0}^{R_{\rm{L}}} \frac{\dd \bar{P}}{\bar{\rho}}.
    \label{eq:phi_thick}
\end{equation}
In the case of Roche-lobe underfilling donor, $\delta R_{\rm d} < 0$, the radius $R_{\rm{L}}$ is above donor's photosphere. Thus, the radius $R_0$ is now defined by
\begin{equation}
    \phi_1 - \phi_0 \equiv \phi_V (R_{\rm{L}}) - \phi_V( R_{\rm d}) + \int_{R_0}^{R_{\rm{d}}} \dd \bar{\phi},
    \label{eq:phi_thin}
\end{equation}
with the help of the potential approximation $\phi_V$ given by equation~(\ref{eq:phi_vol}). 

With values $\rho_0$ and $T_0$ available, we solve the set of 1D hydrodynamic equations~(\ref{eq:hydro_gen}) to obtain $v_0$. In order to reasonably match 1D stellar model to our MT model, $v_0$ has to satisfy $v_0 \ll c_0$ so that the donor's structure is not significantly altered by MT at $R_0$. If $v_0 \approx c_0$, then the donor's profile is significantly different from the hydrostatic one already at $R_0$ which means that the hydrostatic values $\rho_0$ and $T_0$ are not a valid approximation for the true values of $\rho$ and $T$ at the inner boundary of the MT region. This problem arises when we choose the point $R_0$ too close to $R_{\rm{L}}$ so that the difference $\phi_1-\phi_0$ is too small. Conversely, if we choose $R_0$ too far from $R_\text{L}$, the entropy at that point can differ from the regions closer to the surface, which adversely affects the value of $\dot{M}_\text{new}$. Thus, we need to find an optimal $R_0$ with respect to $R_{\rm{L}}$ so that the hydrostatic values $\rho_0$ and $T_0$ are realistic at $\phi_0$ and at the same time the entropy variations do not affect $\dot{M}_\text{new}$ too much. 

The realistic EOS used in this work is adapted from MESA EOS module \citep{paxton2011, paxton2013, paxton2015, paxton2018, paxton2019}. Depending on the metallicity $Z$ and the position in the $\rho T$ plane, one of the following components is used: OPAL/SCVH \citep{rogers2002,saumon1995}, Free EOS \citep{irwin2008},  HELM \citep{timmes2000}, PC \citep{potekhin2010}, Skye \citep{jermyn2021}, or CMS \citep{chabrier2019}. For a given metallicity $Z$ in the donor's photosphere we construct uniform grid of $800 \times 800$ points in the $\log\rho$--$\log T$ plane. The density range depends on the type of donor whose atmosphere we are trying to model, but typically covers 8 dex between $-13 \le \log\rho \le -2$. The temperature range is always $2 \le \log T \le 6$. Using the tutorial by \citet{timmes2021}, we construct a table with all relevant thermodynamic variables evaluated at each grid point. We interpolate in the table using bilinear interpolation \citep[Chapter 3]{press2007}. For $P$ and $\epsilon$ we use both gaseous and radiative components.

\section{Results}\label{sec:results}

In this Section, we show the solutions to the set of equations~(\ref{eq:hydro_gen}). In Section~\ref{sec:res_iso}, we analyze the algebraic solution for isothermal gas. In Section~\ref{sec:res_id}, we provide the same for the case of ideal gas. Finally, in Section~\ref{sec:res_real}, we investigate the solutions using a realistic EOS.

\subsection{Isothermal gas}\label{sec:res_iso}
We were able to find algebraic solution for velocity and density profiles of the set of equations~(\ref{eq:hydro_iso}) as
\begin{subequations}
	\label{eq:hydro_iso_prof}
	\begin{align}
	\frac{1}{2} \left[ \lp \frac{v(x)}{c_T}\rp^2 - \lp \frac{v_0}{c_T}\rp^2 \right] - \ln \frac{v(x)}{v_0} &= - \frac{\phi_{\rm R}^x (x) - \phi_0}{c_T^2}, \label{eq:hydro_iso_v} \\
	\frac{\rho(x)}{\rho_0} &= \frac{v_0}{v(x)},\label{eq:hydro_iso_rho}
	\end{align}
\end{subequations}
where $v_0 \equiv v (x_0)$ is the solution of the algebraic equation
\begin{equation}
    \frac{1}{2} \lp \frac{v_0}{c_T} \rp^2 - \ln \frac{v_0}{c_T} = \frac{\phi_1-\phi_0}{c_T^2} + \frac{1}{2} = \kappa + \frac{1}{2}, 
    \label{eq:hydro_iso_v_0}
\end{equation}
where we define the dimensionless potential difference between the $x_0$ and $x_1$ points as 
\begin{equation}
\kappa \equiv \frac{\phi_1-\phi_0}{c_T^2}.
\end{equation}
This potential difference can be also interpreted as a measure of physical distance between L1 and the starting point of the integration $x_0$.
The MT rate is then given by
\begin{equation}
    - \dot{M}_{\rm d} \equiv \dot{M}_{\rm{thin}} = \frac{2 \pi}{\sqrt{BC}} c_T^2 v_0 \rho_0  = \frac{2 \pi}{\sqrt{BC}} \frac{k}{\overline{m}} v_0 \rho_0 T_0.
    \label{eq:Mdot_iso}
\end{equation}
This MT rate closely resembles the form for optically-thin MT and we will subsequently refer to this value as $\dot{M}_\text{thin}$. We discuss the comparison in more detail in Section~\ref{sec:comp_iso}. 

\begin{figure}
    \centering
	\includegraphics[width=\columnwidth]{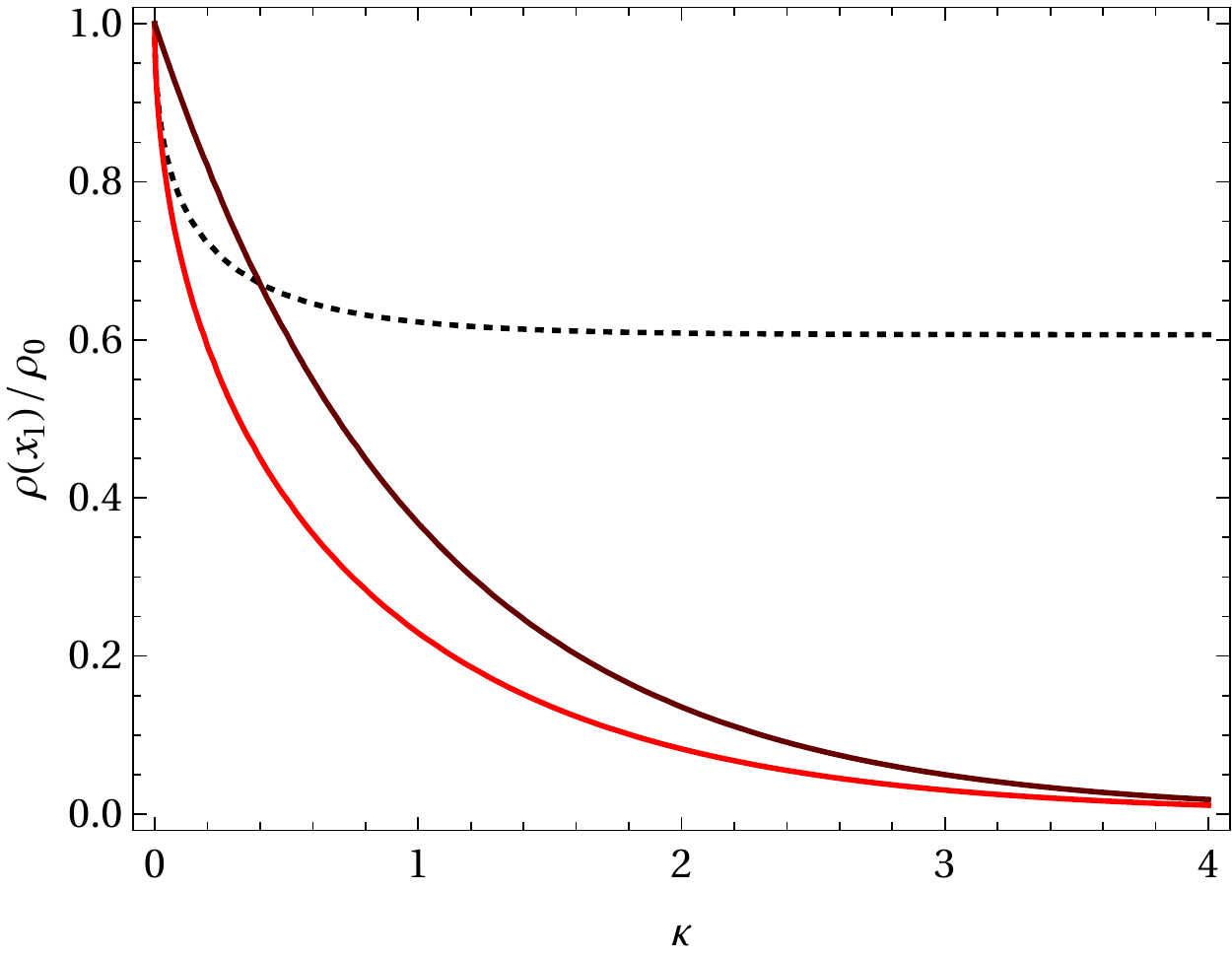}
    \caption{The density $\rho$ at $x_1$ as a function of the dimensionless potential difference $\kappa$ between  $x_0$ and $x_1$ in the case of isothermal gas. The red line shows the density $\rho_1$ as the solution of the set hydrodynamic equations~(\ref{eq:hydro_iso_prof}), which is equivalent to the solution of equation~(\ref{eq:hydro_iso_v_0}) as explained in the text. The dark red line shows the hydrostatic solution $\bar{\rho}_1$ (equation~\ref{eq:iso_rho_stat}). The black dotted line shows $\rho_1/\bar{\rho}_1$, which approaches $1 / \sqrt{e} \doteq 0.61$.}
    \label{fig:iso_rho_1}
\end{figure}

We begin by analyzing the isothermal solution by first looking at the density at L1, $\rho_1$. In Fig.~\ref{fig:iso_rho_1}, we show with the red line the ratio $\rho_1/\rho_0 = v_0 / c_T$ (from equation \ref{eq:hydro_iso_rho}) as a function of the potential difference $\kappa$. We see that when $\kappa$ approaches zero, which means that the distance between $x_0$ and $x_1$ is very small, the two densities are very similar. Conversely, as $\kappa$ gets larger, the density ratio decreases. It is instructive to compare our solution of $\rho_1$ to the isothermal hydrostatic solution,
\begin{equation}
    \frac{\bar{\rho} (x)}{\rho_0} = \exp \lp -\frac{\phi_{\rm R}^x (x) - \phi_0}{c_T^2} \rp, \quad \frac{\bar{\rho}_1}{\rho_0} = \exp (-\kappa),
    \label{eq:iso_rho_stat}
\end{equation}
where we denote $\bar{\rho} \lp x_1\rp = \bar{\rho}_1$, which we show with a dark red line. We see that the hydrodynamic solution is always below the hydrostatic one and that their ratio approaches $1 / \sqrt{e}$ as the potential difference $\kappa$ increases.

\begin{figure}
    \centering
	\includegraphics[width=\columnwidth]{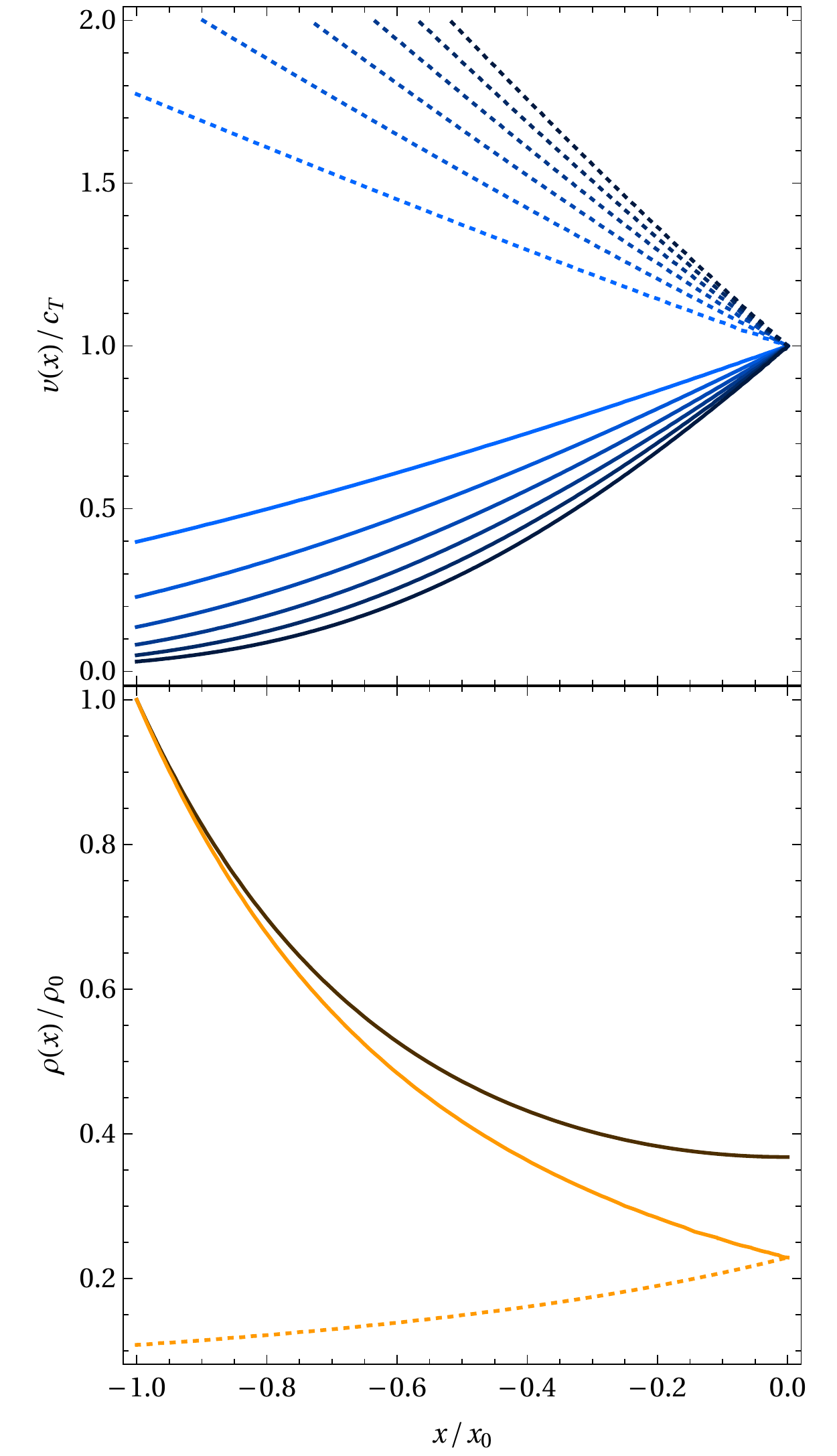}
    \caption{Velocity and density profiles obtained by solving the set of hydrodynamic equations~(\ref{eq:hydro_iso_prof}) for isothermal gas. Physical solutions are indicated by solid lines, non-physical ones by dotted lines. Upper panel shows velocity profiles, where different curves are for different values of $\kappa = 0.5,1.0,\ldots,3.0$. Brighter curves are for lower values of $\kappa$. Lower panel shows the density profile (orange line) for $\kappa = 1$. Solid black line shows the hydrostatic profile  for comparison (equation~\ref{eq:iso_rho_stat}).}
    \label{fig:iso-v_rho_prof}
\end{figure}

Now we proceed to analyze the velocity and density profiles as functions of the $x$ coordinate, which is more intuitive than the potential coordinate $\phi^x_{\rm{R}}$. In order to do that, we need an approximation of the potential coordinate. For simplicity, we use the lowest order approximation from  equation~(\ref{eq:phi_app}), $\phi_{\rm R}^x (x) = - Ax^2 / 2 +\phi_1$, which provides the relation between physical and potential coordinates, 
\begin{equation}
    \frac{x}{x_0} = - \sqrt{\frac{\phi_1-\phi_{\rm{R}}^x (x)}{\phi_1-\phi_0}}.
    \label{eq:x_coor}
\end{equation}
In the upper panel of Fig.~\ref{fig:iso-v_rho_prof}, we show the velocity profiles for different values of $\kappa$. We see that there are two branches of solutions satisfying the critical boundary condition $v(x_1) = c_T$. One class of solutions is subsonic (solid lines), while the other one is supersonic (dotted lines). Only the subsonic solutions are physically relevant because the gas has to start with negligible velocity at the donor's surface. Multiple branches of solutions are common in two-point boundary value problems ranging from stellar winds to accretion onto proto-neutron stars \citep[e.g.,][]{lamers1999,yamasaki05,pejcha12}.

In the lower panel of Fig.~\ref{fig:iso-v_rho_prof}, we show the density profile for $\kappa =1$ along with the corresponding hydrostatic density profile. The part of the density profile that grows with $x$ (dotted line) corresponds to the non-physical supersonic velocity profile. The hydrodynamic solution is always below the hydrostatic one and reaches roughly $1 / \sqrt{e}$ times the hydrostatic value at the $x_1$ point, as we already showed in Fig.~\ref{fig:iso_rho_1}.

\subsection{Ideal gas}\label{sec:res_id}
We show the process of solving our model for ideal gas  in Appendix~\ref{sec:hydro_id}. The solution can be written as a set of algebraic equations,
\begin{subequations}
	\label{eq:id_sol}
	\begin{align}
	    \frac{1}{2} \left[ \lp \frac{v}{c_0}\rp^2 - \lp\frac{v_0}{c_0}\rp^2 \right] + \frac{\Gamma}{\Gamma-1} \left[\lp \frac{v_0}{v}\rp^{\frac{\Gamma-1}{\Gamma}} - 1\right]  &= - \frac{\phi_{\rm R}^x - \phi_0}{c_0^2}, \label{eq:id_v} \\
	    \rho  &= \lp \frac{v_0}{v }\rp^{\frac{1}{\Gamma}} \rho_0, \label{eq:id_rho} \\
        T  &= \lp\frac{v_0}{v}\rp^\frac{\Gamma-1}{\Gamma} T_0, \label{eq:id_T}
	\end{align}
\end{subequations}
where $c_0^2 \equiv c_T^2 (x_0) = k T_0 / \overline{m}$ and $v_0$ is the solution of
\begin{equation}
    \frac{3\Gamma-1}{2(\Gamma-1)} \lp \frac{v_0}{c_0}\rp^{\frac{2(\Gamma-1)}{3\Gamma-1}} - \frac{1}{2} \lp \frac{v_0}{c_0}\rp^2 = - \frac{\phi_1-\phi_0}{c_0^2} + \frac{\Gamma}{\Gamma-1} = \kappa_{\rm{max}} - \kappa.
    \label{eq:id_v_0}
\end{equation}
Here, we denote $\kappa_\text{max} \equiv \Gamma / (\Gamma-1)$ and $\kappa \equiv (\phi_1-\phi_0) / c_0^2$. If $\kappa > \kappa_{\rm{max}}$, then equation~(\ref{eq:id_v_0}) does not have a physical solution. Physically, this means that the L1 point lies above the point, where the density drops to zero. The case $\kappa=\kappa_{\rm{max}}$ is the limiting case when $\rho (x_1) = 0$, $T(x_1) = 0$ and thus $v(x_1) = 0$. The MT rate through L1 in the case of ideal gas is given by
\begin{equation}
    - \dot{M}_{\rm d} \equiv \dot{M}_{\rm{thick}} = \frac{2 \pi}{\sqrt{BC}} c_0^2 v_0 \rho_0 = \frac{2 \pi}{\sqrt{BC}} \frac{k}{\overline{m}} v_0 \rho_0 T_0.
    \label{eq:Mdot_id}
\end{equation}
Since the flow is adiabatic, we expect that this MT rate will be applicable to the optically-thick case and we will subsequently refer to this value as $\dot{M}_\text{thick}$. We note that the equations for ideal gas reduce to isothermal gas if we perform the limit $\Gamma \rightarrow 1$ and use the identity $\lim_{x \to 0} (a^x-1)/x = \ln a$, which is valid for $a>0$.

\begin{figure}
    \centering
	\includegraphics[width=\columnwidth]{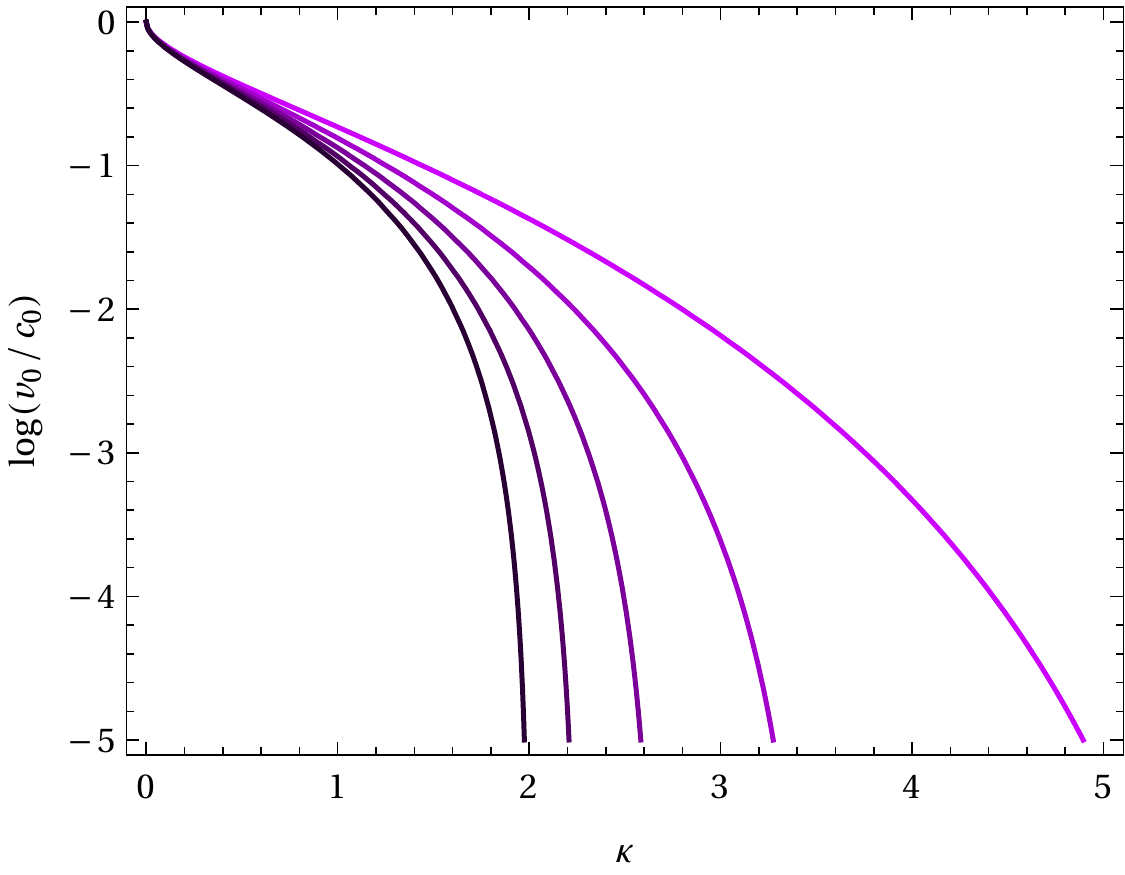}
    \caption{Solutions for $v_0$ for ideal gas given by equation~(\ref{eq:id_v_0}). Lines show solutions for different values of $\Gamma = 1.2, 1.4, \ldots, 2.0$, where brighter curves indicate lower values of $\Gamma$.}
    \label{fig:id_v0}
\end{figure}

\begin{figure}
    \centering
	\includegraphics[width=0.98\columnwidth]{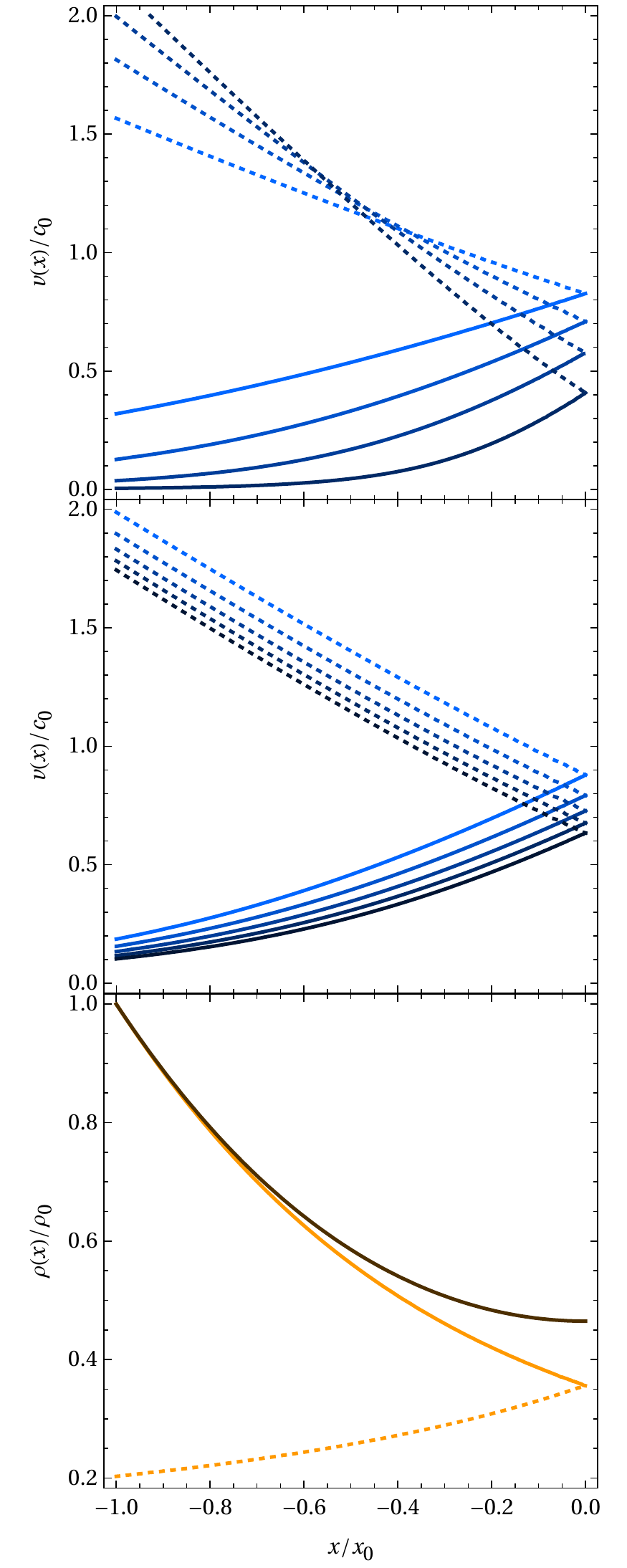}
    \caption{Velocity and density profiles for ideal gas. Physical solutions are indicated by solid lines while non-physical solutions by dotted lines. Top panel shows velocity profiles for $\Gamma=5/3$ and $\kappa =0.5,1.0,\ldots,2.0$. Brighter curves indicate lower values of $\kappa$. Middle panel shows velocity profiles for $\kappa  = 1$ and $\Gamma = 1.2,1.4,\ldots,2.0$. Brighter curves indicate lower values of $\Gamma$. Bottom panel shows the density profile (orange line) for $\Gamma=5/3$ and $\kappa = 1$. The hydrostatic profile is shown with a dark orange line.}
    \label{fig:id-v_rho_prof}
\end{figure}

In Fig.~\ref{fig:id_v0}, we show the solutions for $v_0$ as a function of $\kappa$ and for different values of $\Gamma$.
We see that as $\Gamma$ decreases $\kappa_{\rm{max}}$ increases. This is because for lower values of $\Gamma$ the EOS is less stiff and the point where the density drops to zero occurs further away.
In the top and middle panels of Fig.~\ref{fig:id-v_rho_prof}, we show the velocity profiles $v(x)$ for various values of $\Gamma$ and $\kappa$. We see that the critical velocity $v(x_1)$ reached at L1 varies with both $\Gamma$ and $\kappa$. With decreasing $\kappa$ and decreasing potential difference between $x_0$ and $x_1$, the critical velocity $v(x_1)$ increases. At the same time, for lower values of $\Gamma$ the critical velocity $v (x_1)$ increases. Similarly to the isothermal solution, there are two branches of solutions, where the one with decreasing supersonic velocity is non-physical (dotted lines).

\begin{figure}
    \centering
	\includegraphics[width=\columnwidth]{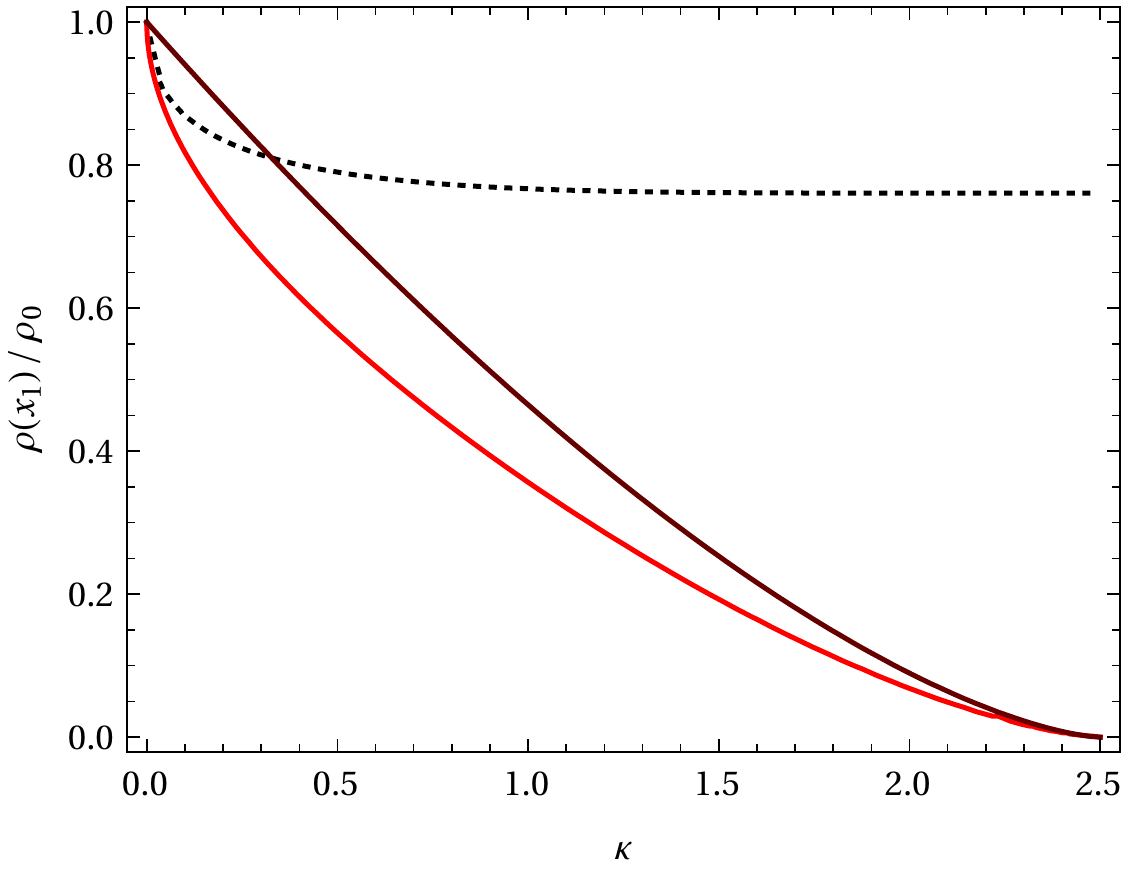}
    \caption{The density $\rho$ at $x_1$ as a function of $\kappa$ for ideal gas with $\Gamma=5/3$. The red line shows the density $\rho_1$ as the solution of equations~(\ref{eq:id_sol}). The dark red line shows the hydrostatic solution $\bar{\rho}_1$ (equation~\ref{eq:id_rho_stat}). The black dotted line shows $\rho_1/\bar{\rho}_1$, which is approaching $(5/6)^{3/2} \doteq 0.76$.}
    \label{fig:id_rho1}
\end{figure}

Now we turn to discuss the density profile. In the bottom panel of Fig.~\ref{fig:id-v_rho_prof}, we show the density profile for $\Gamma = 5/3$ and $\kappa =1$. In Fig.~\ref{fig:id_rho1}, we show the density at L1 point $\rho(x_1)$ as a function of $\kappa$. It is instructive to compare the density profile to a hydrostatic profile. Since our MT model with ideal gas is adiabatic by construction, it is possible to use the the polytropic approximation (equation~\ref{eq:polytrope}) to write the hydrostatic profile as
\begin{equation}
    \frac{\bar{\rho} (x)}{\rho_0} = \lp 1 - \frac{1}{\kappa_{\rm{max}}}\frac{\phi_{\rm R}^x (x) - \phi_0}{c_0^2} \rp^{\frac{1}{\Gamma-1}}, \quad \frac{\bar{\rho}_1}{\rho_0} = \lp 1 - \frac{\kappa}{\kappa_{\rm{max}}}\rp^{\frac{1}{\Gamma-1}}.
    \label{eq:id_rho_stat}
\end{equation}
We again see that the hydrostatic density is always higher than the hydrodynamic one, as expected in a moving medium. The ratio of the hydrodynamic over the hydrostatic density approaches $\rho(x_1)/\bar{\rho}(x_1) \rightarrow [2\Gamma / (3\Gamma-1 )]^{1 / (\Gamma-1)}$ as $\kappa \rightarrow \kappa_\text{max}$. For $\Gamma=5/3$ we have $\kappa_{\rm{max}} = 5 / 2$ and  $\rho(x_1)/\bar{\rho}(x_1) \rightarrow ( 5 / 6 )^{3 / 2} \doteq 0.76$, as we illustrate in Fig.~\ref{fig:id_rho1}.

\subsection{Realistic equation of state}
\label{sec:res_real}

\begin{figure}
    \centering
	\includegraphics[width=\columnwidth]{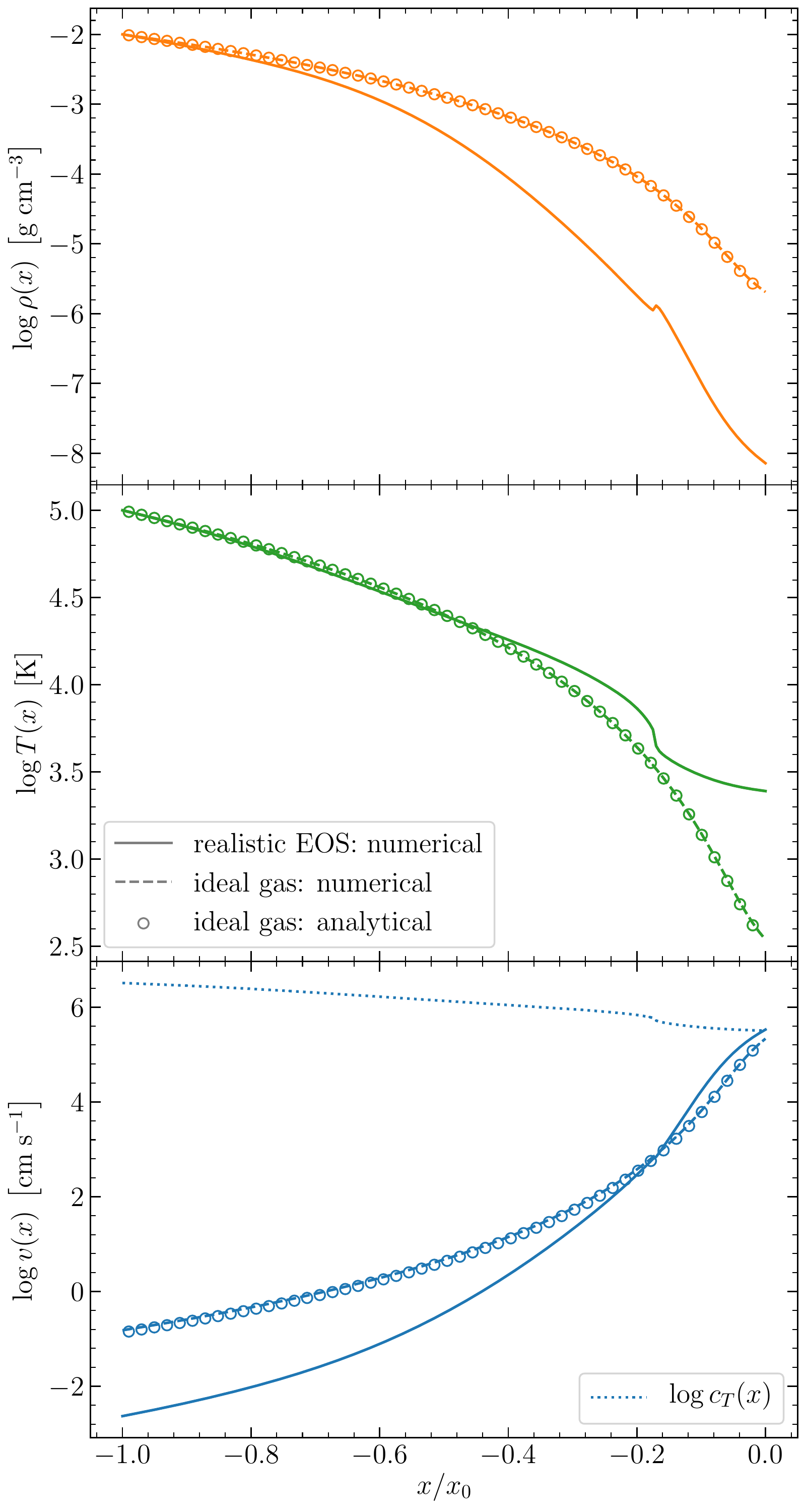}
    \caption{Density $\rho$, temperature $T$, and velocity $v$ profiles for realistic EOS (solid line). For comparison, solutions with the same initial conditions but with an ideal gas are displayed for $\kappa = 2.49$ and $\Gamma = 5/3$. We show algebraic solution with dashed line and numerical solution with our relaxation code with empty circles. The dotted line in the bottom panel marks the value of the isothermal sound speed $c_T(x)$, $v(x_1) = c_T(x_1)$. }
    \label{fig:real_v_id-prof}
\end{figure}

We first discuss features of our solution with realistic EOS for one set of parameters (Sec.~\ref{sec:real_prop}). Next, we illustrate how to calculate MT rate based on an external 1D hydrostatic stellar model (Sec.~\ref{sec:real_hydrostatic}) and then we calculate the MT rate for a solar-type star on the main sequence (Sec.~\ref{sec:sun-like-ms_R0}), for a red giant (Sec.~\ref{sec:sun-like-rgb_R0}), and for a massive star undergoing thermal time-scale MT (Sec.~\ref{sec:30-Msun_R0}).

\subsubsection{Properties of the solution}
\label{sec:real_prop}
In Fig.~\ref{fig:real_v_id-prof}, we present hydrodynamic profiles of $\rho$, $T$, and $v$ for realistic EOS. We choose metallicity $Z = 0.02$ and the relative metal fractions following the MESA \texttt{1M\_pre\_ms\_to\_wd} test suite case. The boundary conditions are $\rho_0 = 10^{-2}$ g cm$^{-3}$ and $T_0 = 10^5$ K. Additionally, we show solutions with the same boundary conditions at $x_0$ but for an ideal gas. We set $\kappa= 2.49$, $\overline{m} = \mu m_{\rm{u}}$, where $\mu = 0.617$, and $\Gamma = 5/3$, which gives $\kappa_{\rm{max}} = 2.5$. The solution for ideal gas can be obtained either from algebraic equations (dotted line) or using our numerical relaxation code (open points). There are two observations that we can make here. First, the two solution methods for ideal gas agree very well even for our extreme choice of $(\kappa_{\rm{max}} - \kappa) / \kappa_\text{max} = 0.004$, which validates our numerical scheme. Second, the realistic temperature profile shows a steep temperature drop around $4000$\,K, which arises due to hydrogen recombination.  The small jump in $\rho$ at $x/x_0\approx -0.19$ is due to a very rapid change in $\Gamma$, which occurs at a region of $\rho$ and $T$, where the MESA EOS blends together FreeEOS and OPAL/SCVH. We verified that the mass flux is conserved to a high precision across the entire range of $x$. The densities are significantly lower for realistic EOS compared to the ideal gas, but the velocities at L1 are very similar. This implies that MT rates obtained for the two EOSs will be very different. We cannot compute the absolute MT rate without binary parameters $B$ and $C$, but we can calculate their ratio. We find $\dot{M}_{\rm{thick}} / \dot{M}_{\rm{new}} \approx 10^2$, which indicates that the realistic EOS is an important part of our model.

\subsubsection{Relation to hydrostatic stellar structure}
\label{sec:real_hydrostatic}
In order to connect our MT model with 1D hydrostatic stellar structure, we need to set the relative radius excess $\delta R_\text{d}$ and the inner point of our integration $R_0$. It is useful to express the distance between $R_0$ and $R_{\rm{L}}$ (or $x_0$ and $x_1$) with the number of pressure scale heights $\Delta N_{H_P}$. We define the pressure scale height number $N_{H_P}$ for the donor's interior as
\begin{equation}
    N_{H_P} (R) \equiv \int_{R}^{R_{\rm{d}}} \frac{\dd R^\prime}{H_P (R^\prime)}, \quad \rm{for} \quad  R < R_\dd,
\end{equation}
and for the exterior as 
\begin{equation}
    N_{H_P} (R) \equiv - \frac{\phi_V (R) - \phi_V (R_{\rm{d}}) }{P_{\rm{ph}} / \rho_{\rm{ph}}}, \quad \rm{for} \quad R > R_\dd,
\end{equation}
where $H_P$ is the pressure scale height and $P_\ph$ is the pressure at the photosphere. We see that $N_{H_P} (R_\dd) = 0$ and that $N_{H_P}$increases inward and decreases outward. Hence, for a Roche-lobe overflowing donor we have
\begin{equation}
    \Delta N_{H_P} = N_{H_P} (R_0) - N_{H_P} (R_{\rm{L}}) = \int_{R_0}^{R_{\rm{L}}} \frac{\dd R}{H_P}.
    \label{eq:DN_HP_thick}
\end{equation}
In the case of Roche-lobe underfilling donor we have
\begin{equation}
    \begin{aligned}
        \Delta N_{H_P} &= N_{H_P} (R_0) - N_{H_P} (R_{\rm{d}}) + N_{H_P} (R_{\rm{d}}) - N_{H_P} (R_{\rm{L}}) \\ 
        &= \int_{R_0}^{R_{\rm{d}}} \frac{\dd R}{H_P} + \frac{\phi_V (R_{\rm{L}}) - \phi_V ( R_{\rm{d}}) }{P_{\rm{ph}} / \rho_{\rm{ph}}}.
         \label{eq:DN_HP_thin}
    \end{aligned}
\end{equation}
By looking at the dependencies $\dot{M}_{\rm{new}} (R_0)$ we can determine an optimal position of $R_0$ in units of $\Delta N_{H_P}$ for a given type of donor.

\subsubsection{$1\,\Msun$ donor on the main sequence}
\label{sec:sun-like-ms_R0}

\begin{figure}
    \centering
	\includegraphics[width=\columnwidth]{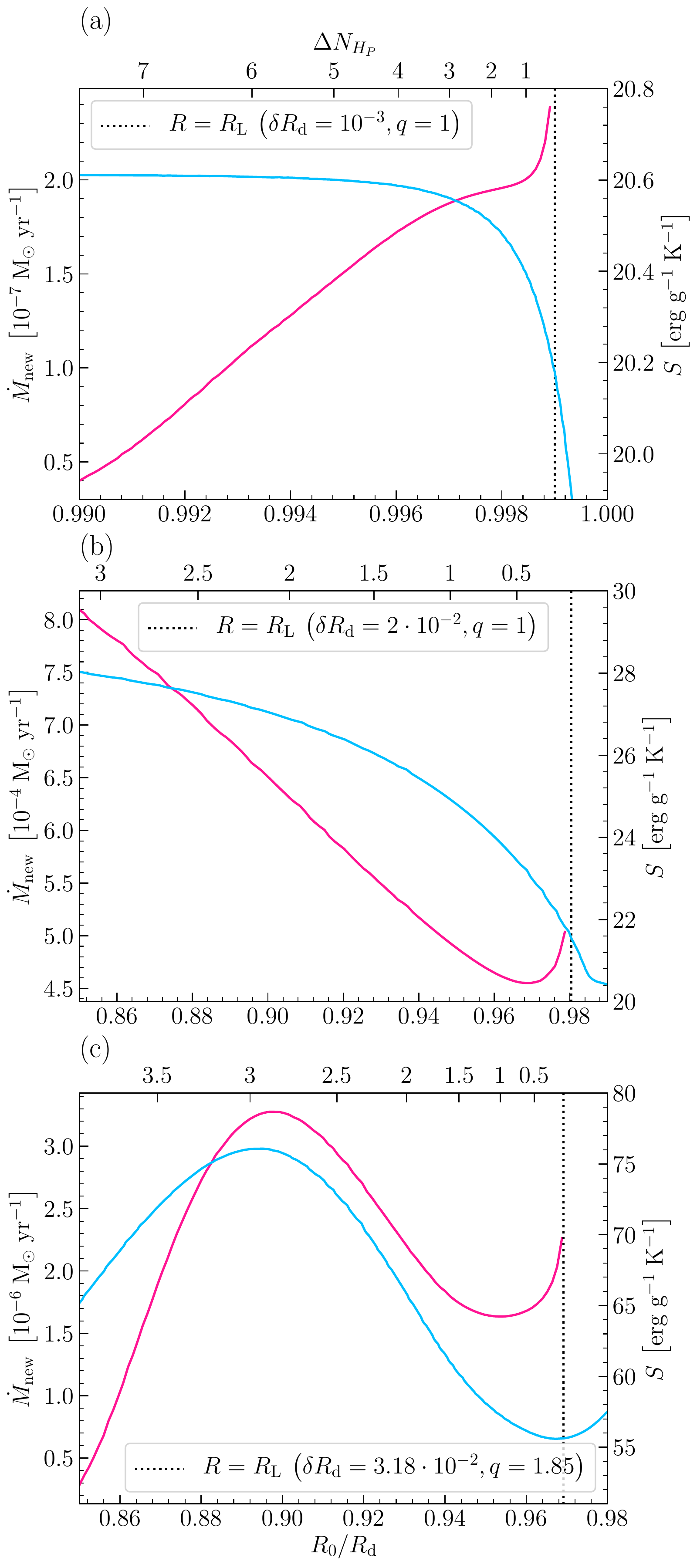}
    \caption{The dependency of $\dot{M}_\new$ on $R_0$ (pink lines) for a given constant value of $\delta R_\dd$ and the corresponding profile of entropy per gram $S$ (blue lines). The top horizontal axis in all panels shows the number of pressure scale heights $\Delta N_{H_P}$ between $R_0$ and $R_{\rm{L}}$ (equation~\ref{eq:DN_HP_thick}). We show three cases: (a) $1\,\Msun$ donor on the main sequence, (b) $1\,\Msun$ donor on the red giant branch, and (c) $30\,\Msun$ low-metallicity star undergoing thermal  MT.}
    \label{fig:Mdot-R0}
\end{figure}

We use MESA version r21.12.1 and its \texttt{1M\_pre\_ms\_to\_wd} test suite case to compute hydrostatic profiles of a donor star with initial mass $1\,\Msun$ and metallicity $Z = 0.02$. To obtain profiles representative for a main sequence star, we evolve the model to $2.8$ Gyr, when $T_\text{eff} = 5.7 \times 10^3$ K, $R_\text{d} = 0.95\,\Rsun$, and luminosity is $L = 0.86\,\Lsun$. In the top panel of Fig.~\ref{fig:Mdot-R0}, we show $\dot{M}_{\rm{new}}(R_0)$ for $\delta R_{\rm{d}} = 10^{-3}$ and $q = 1$ along with the entropy profile of the stellar model. We see that $\dot{M}_{\rm{new}}$ varies with $R_0$. For $R_0$ close to $R_{\rm{L}}$, $\dot{M}_{\rm{new}}$ increases because taking the hydrostatic values $\bar{\rho} (R_0)$ and $\bar{T} (R_0)$ as the initial values $\rho_0$ and $T_0$ overestimates the true hydrodynamic values, and consequently also $\dot{M}_\text{new}$. For $R_0$ far away from $R_{\rm{L}}$, the MT rate decreases mainly because the true density and temperature profiles are not adiabatic over the studied range. This leads to an optimal value of $R_0$, which has to lie between these two extreme regimes. By investigating the dependence $\dot{M}_{\rm{new}} (R_0)$ for different values of $\delta R_\dd$ we conclude that the optimal placement of $R_0$ for this donor is around $\Delta N_{H_P} = 2.0$--$3.0$. For the specific case of $\delta R_\dd = 10^{-3}$ and $q=1$ this choice gives $\dot{M}_{\rm{new}} \approx 2 \times 10^{-7}\,\Msun\,\rm{yr}^{-1}$.

\subsubsection{$1\,\Msun$ donor on the red giant branch}\label{sec:sun-like-rgb_R0}
We evolve the stellar model further to age $12.3$\,Gyr, when we obtain a red giant with $T_\text{eff} = 3.4 \times 10^3$\,K, $R_\text{d} = 89\,\Rsun$, and $L = 1.0 \times 10^3\,\Lsun$. In the middle panel of Fig.~\ref{fig:Mdot-R0}, we show $\dot{M}_\new (R_0)$ for $\delta R_\dd = 2 \times 10^{-2}$ and $q=1$.  We see that $\dot{M}_\text{new}$ depends on $R_0$, because the stellar profile is not adiabatic. It appears that $\dot{M}_\text{new}$ and the entropy are correlated. By investigating $\dot{M}_{\rm{new}} (R_0)$ for various values of $\delta R_\dd$, we conclude that the optimal position of $R_0$ for this donor is around $\Delta N_{H_P} = 0.5$--$1.0$. For the specific case of $\delta R_\dd = 2 \times 10^{-2}$ and $q=1$ we obtain $\dot{M}_{\rm{new}} \approx 4.6 \times 10^{-4}\,\Msun\,\rm{yr}^{-1}$.

\subsubsection{30 $\Msun$ donor undergoing thermal time-scale mass transfer}\label{sec:30-Msun_R0}
We use $30\,\Msun$ low-metallicity donor undergoing intensive thermal time-scale MT to a black hole, which was investigated by \citet{marchant2021}. The donor's initial metallicity is $Z = \Zsun/10$, where $\Zsun = 0.0142$ \citep{asplund2009} and where the relative metal mass fractions are from \citet{grevesse1998}. Out of the many black hole masses and initial orbital periods studied by \citet{marchant2021}, we choose the model with a  $7.5\,\Msun$ black hole and an initial period of $31.6$ days. We rerun this case using the full MT prescription $\dot{M}_{\rm{M}}$ of \citet{marchant2021} using the same MESA version and we selected a model near the end of the thermal time-scale MT phase with age $6.79\,\rm{Myr}$, $M_\text{d} = 14.0\,\Msun$, core-helium abundance $0.36$, $\delta R_\text{d} = 3.18 \times10^{-2}$, and $q = 1.85$.  In the bottom panel of Fig.~\ref{fig:Mdot-R0}, we show $\dot{M}_\new (R_0)$ and entropy of the underlying stellar model. We see that $\dot{M}_\text{new}$ shows a complex behavior as a function of $R_0$, which is closely correlated with variations in the entropy. By investigating $\dot{M}_{\rm{new}} (R_0)$ for different stages of the binary evolution we conclude that the optimal position of $R_0$ is at around $\Delta N_{H_P} = 0.5$--$1.0$. For this specific case, this choice leads to $\dot{M}_{\rm{new}} \approx 1.7 \times 10^{-6}\,\Msun\,\rm{yr}^{-1}$.

\section{Comparison with existing models}\label{sec:comparison}

In this section, we compare our new model with the existing models for the cases of isothermal gas (Sec.~\ref{sec:comp_iso}), ideal gas (Sec.~\ref{sec:comp_ideal}), and realistic EOS (Sec.~\ref{sec:comp_real}). In Section~\ref{sec:comp_real}, we also estimate the effects of our new MT model on the evolution of the binary.

\subsection{Isothermal gas}
\label{sec:comp_iso}

\begin{figure}
    \centering
	\includegraphics[width=\columnwidth]{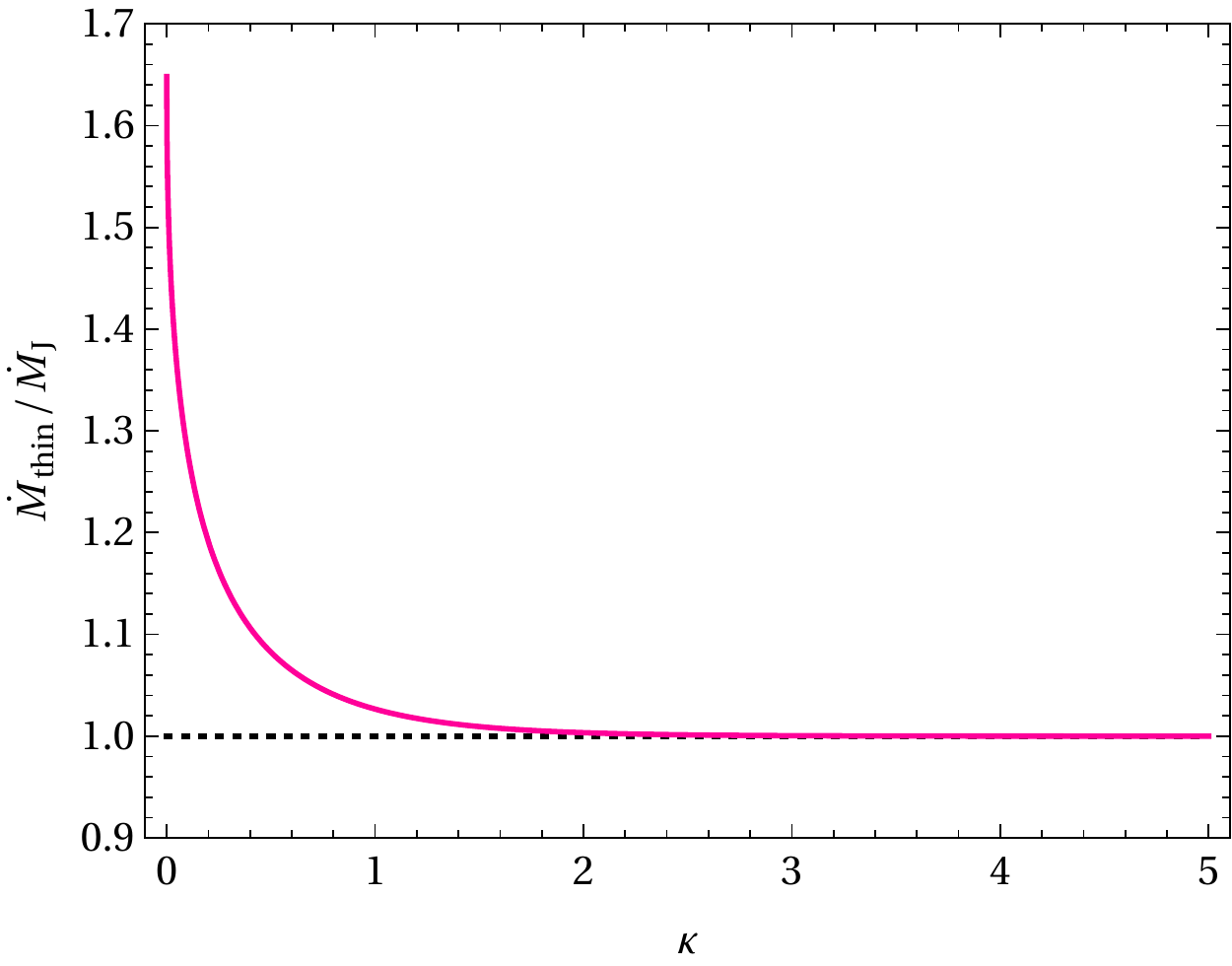}
    \caption{Comparison of our isothermal MT rate $\dot{M}_{\rm{thin}}$ with the MT rate $\dot{M}_{\rm{J}}$ of \citet{jackson2017}.}
    \label{fig:iso_comp}
\end{figure}

Usually, the assumption of isothermal gas flow corresponds to the optically thin MT and we compare our expression for $\dot{M}_\text{thin}$ (equation~\ref{eq:Mdot_iso}) with $\dot{M}_\text{J}$ of \citet{jackson2017} (equation~\ref{eq:Mdot_J}). In this simple case, it is not necessary to reconstruct the donor's surface layers with our model and we choose to identify donor's photosphere with the $x_0$ point, $R_0 = R_{\rm d}$. We can derive an expression for the ratio of MT rates as
\begin{equation}
    \frac{\dot{M}_{\rm{thin}}}{\dot{M}_{\rm J}} = \exp \left[ \frac{1}{2} \lp \frac{v_0}{c_T} \rp^2 \right],
    \label{eq:iso_Mdot_comp}
\end{equation}
where $v_0$ is given by the equation~(\ref{eq:hydro_iso_v_0}). In Fig.~\ref{fig:iso_comp}, we show this ratio as a function of $\kappa$. The ratio starts at $\sqrt{e} \doteq 1.65$ for no potential difference $\kappa = 0$ and approaches $1$ for large $\kappa$. We can explain this result by realizing that for low values of $\kappa$ our model starts at $x_0$ with already large values of velocity, $v_0 \sim c_T$ (see the upper panel of Fig.~\ref{fig:iso-v_rho_prof}), which causes the hydrostatic density $\rho_0$ at $x_0$ to overestimate the true hydrodynamic density at $x_0$. In other words, our choice of $R_0 = R_\text{d}$ causes $\dot{M}_{\rm{thin}}$ to overestimate the value of MT rate when $R_0$ is located very close L1, $\kappa \lesssim 2$.

\subsection{Ideal gas}
\label{sec:comp_ideal}

\begin{figure}
    \centering
	\includegraphics[width=\columnwidth]{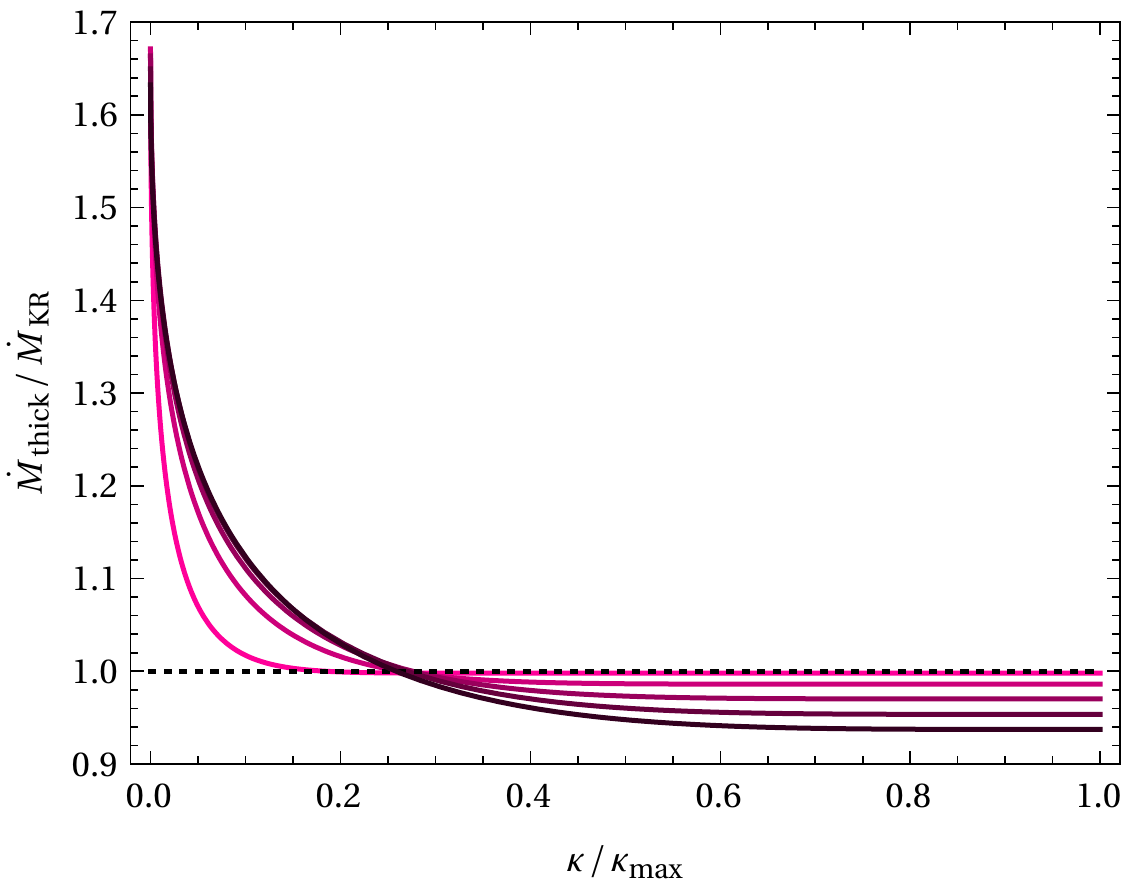}
    \caption{Comparison of our MT rate $\dot{M}_{\rm{thick}}$ with the standard optically-thick MT rate $\dot{M}_{\rm{KR}}$ \citep{kolb1990} in the case of adiabatic ideal gas and polytropic donor. The individual lines are for different values of $\Gamma = 1.1,1.3,\ldots,1.9$ with brighter curves corresponding to lower $\Gamma$.}
    \label{fig:id_comp}
\end{figure}

\begin{figure}
    \centering
	\includegraphics[width=\columnwidth]{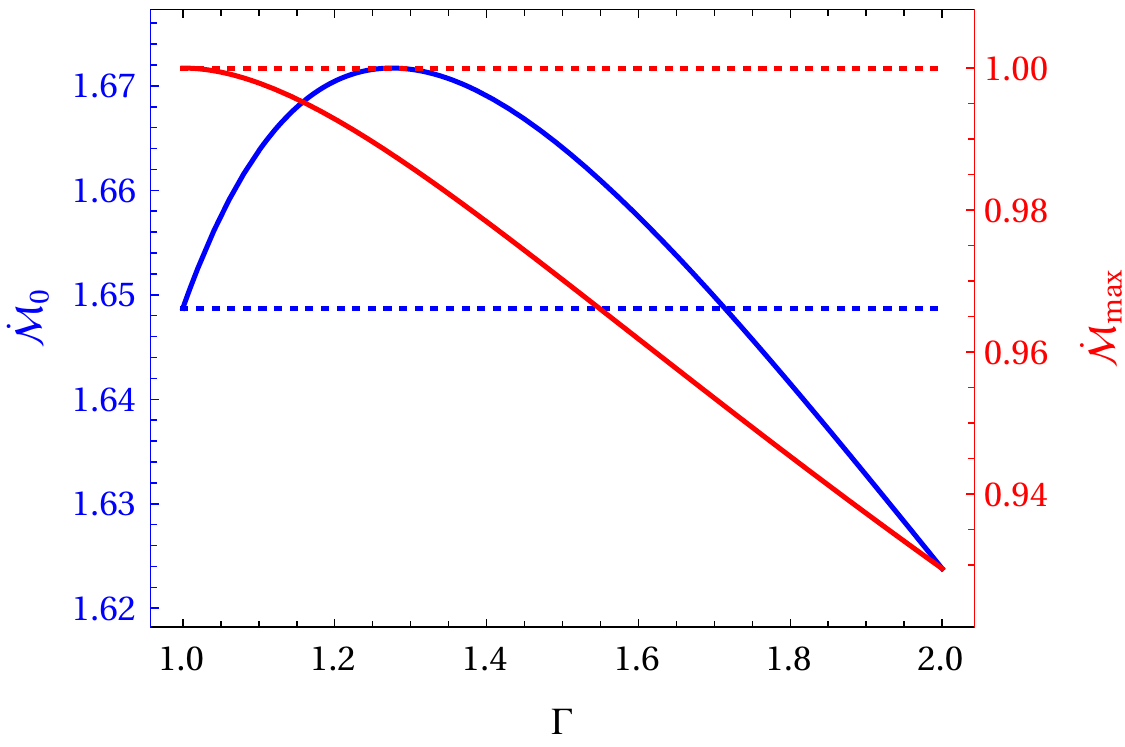}
    \caption{The limiting values $\dot{\mathcal{M}}_{\rm{0}}$ (solid blue line) and $\dot{\mathcal{M}}_{\rm{max}}$ (solid red line) as a function of $\Gamma$. To better illustrate the trends, we also show the isothermal limits (blue and red dotted lines).}
    \label{fig:id_comp_lim_max}
\end{figure}

Since our MT model is adiabatic, the assumption of ideal gas implies that the gas can be described by a polytrope (equation~\ref{eq:polytrope}). Therefore, we consider a polytropic donor with the same polytropic index $\Gamma$ and $\delta R_\text{d}>0$. In such a simple case, the photospheric density of a polytropic donor is $\rho_\ph = 0$, which implies that the saturated optically-thin part of the MT rate is $\dot{M}_{\rm{J,0}}=0$.
In Appendix~\ref{sec:id_Mdot_comp_app}, we calculate the ratio of our MT rate $\dot{M}_\text{thick}$ (equation~\ref{eq:Mdot_id}) to the standard optically-thick MT rate $\dot{M}_\text{KR}$ (equation~\ref{eq:Mdot_KR}). The final result is
\begin{equation}
    \frac{\dot{M}_{\rm{thick}}}{\dot{M}_{\rm{KR}}} = \frac{3 \Gamma-1}{2 \Gamma F_3 } \lp \frac{\rho_0}{\bar{\rho}_1}\rp^{\frac{3\Gamma-1}{2}} \frac{v_0}{c_0} = \frac{3 \Gamma-1}{2 \Gamma F_3 } \lp \frac{\rho_1}{\bar{\rho}_1}\rp^{\frac{3\Gamma-1}{2}}.
    \label{eq:id_Mdot_comp}
\end{equation}
In Fig.~\ref{fig:id_comp}, we show this ratio as function of $\kappa / \kappa_{\rm{max}}$ for different values of $\Gamma$. Similarly to the optically-thin case, the ratio is large for $\kappa = 0$ and approaches $\approx 1$ for $\kappa \rightarrow \kappa_{\rm{max}}$ depending on the value of $\Gamma$. In Appendix~\ref{sec:id_Mdot_comp_app}, we calculate the extreme values of the ratio to obtain
\begin{equation}
    \dot{\mathcal{M}}_{\rm{0}} \equiv \lim_{\kappa\to 0^+} \frac{\dot{M}_{\rm{thick}}}{\dot{M}_{\rm{KR}}} = \frac{3 \Gamma-1}{2 \Gamma F_3 } = \frac{3\Gamma-1}{2 \Gamma^{\frac{3}{2}}} \lp \frac{\Gamma+1}{2}\rp^{\frac{\Gamma+1}{2 (\Gamma-1)}},
    \label{eq:id_Mdot_comp_lim_0}
\end{equation}
and
\begin{equation}
    \dot{\mathcal{M}}_{\rm{max}} \equiv \lim_{\kappa\to\kappa_{\rm{max}}^-} \frac{\dot{M}_{\rm{thick}}}{\dot{M}_{\rm{KR}}} = \Gamma^{-\frac{1}{2}} \left[\frac{\Gamma (\Gamma+1)}{3\Gamma-1}\right]^{\frac{\Gamma+1}{2(\Gamma-1)}}.
    \label{eq:id_Mdot_comp_lim_max}
\end{equation}
We note that results for $\kappa \rightarrow 0$ are not physically relevant, because the inner boundary of our region is too close to L1. The other extreme, $\kappa \rightarrow \kappa_\text{max}$ is more important for understanding the differences between the models. 
 In Fig.~\ref{fig:id_comp_lim_max}, we show the functions $\dot{\mathcal{M}}_{\rm{0}}$ and $\dot{\mathcal{M}}_{\rm{max}}$. We see that these functions are very slowly varying across the range of astrophysically plausible values of $\Gamma$. This implies that the two MT rates  $\dot{M}_\text{thick}$ and $\dot{M}_\text{KR}$ give similar, but not identical values for a wide range of parameters. As we show in Appendix~\ref{sec:id_Mdot_comp_app}, $\lim_{\Gamma \to 1} \dot{\mathcal{M}}_0 = \sqrt{e}$ and $\lim_{\Gamma \to 1} \dot{\mathcal{M}}_{\rm{max}} = 1$, which means that the our optically-thick MT model reduces to the optically-thin case in the limit of $\Gamma \rightarrow 1$.

\subsection{Realistic equation of state}
\label{sec:comp_real}
We first discuss the $1\,\Msun$ donor on the main sequence (Sec.~\ref{sec:comp_ms}) and the red giant branch (Sec.~\ref{sec:comp_rgb}). We then turn to the $30\,\Msun$ donor undergoing thermal time-scale mass transfer (Sec.~\ref{sec:30-Msun_comp}), where we also try to incorporate the effects of our new model on the evolution (Sec.~\ref{sec:comp_evol}). Finally, we discuss the ambiguity in choosing the right matching potential to hydrostatic models (Sec.~\ref{sec:potential_ambiguity}).

\subsubsection{Sun-like donor on the main sequence}
\label{sec:comp_ms}

\begin{figure*}
    \centering
	\includegraphics[width=\textwidth]{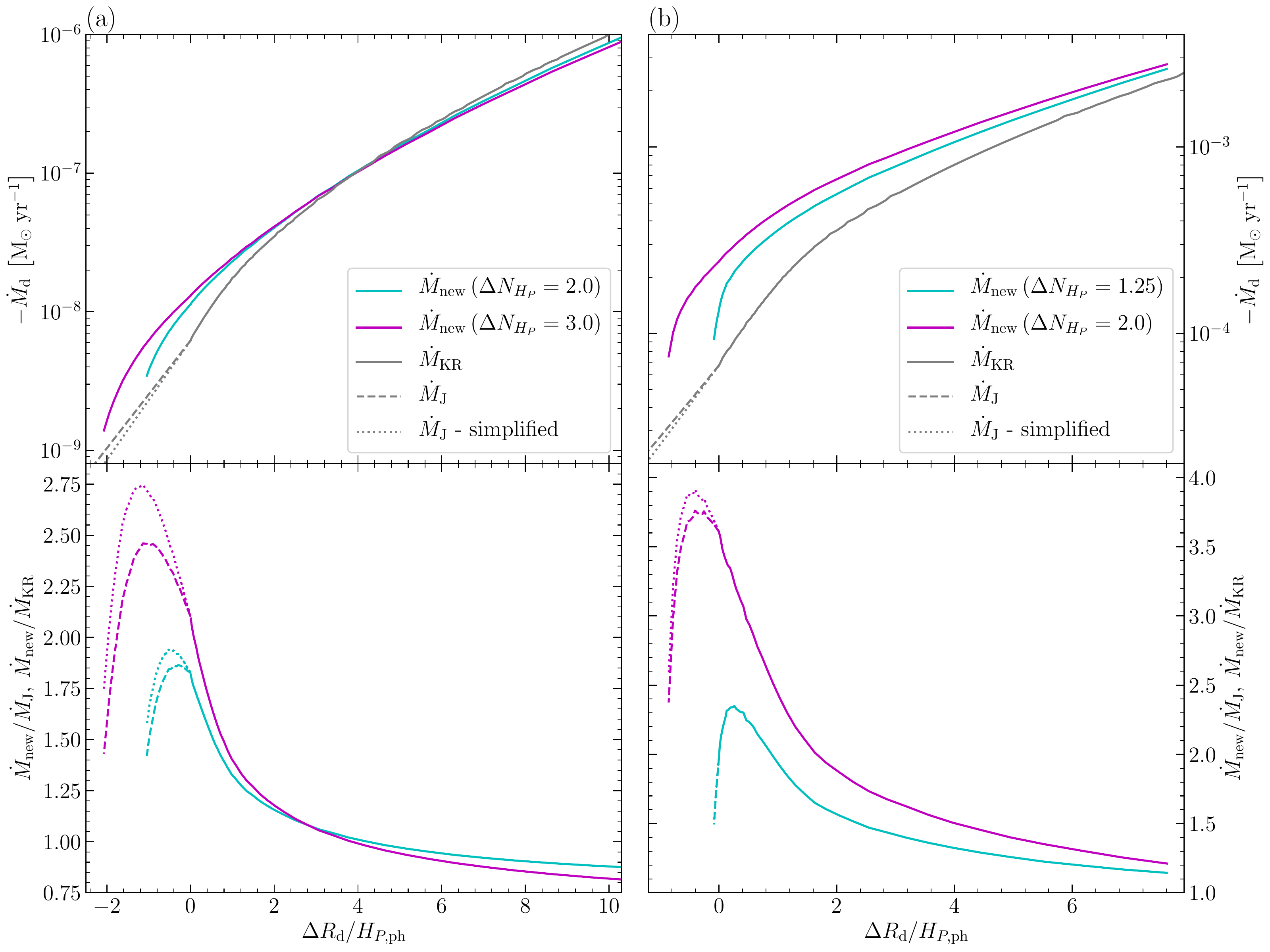}
    \caption{Comparison of our MT rate $\dot{M}_\new$ (magenta and cyan lines) with existing models that combine optically-thin and optically-thick parts (grey lines). We additionally show a modification of of the existing optically-thin model $\dot{M}_\text{J}$, where we apply a simplified version of the potential difference, see Sec. \ref{sec:potential_ambiguity}. Column (a) is for $1\,\Msun$ donor on the main sequence and column (b) is for a red giant. Binary mass ratio is $q = 1$ in both cases.}
    \label{fig:Mdot-DR-1M}
\end{figure*}

In the left column of Fig.~\ref{fig:Mdot-DR-1M}, we compare $\dot{M}_\new (\Delta R_\dd)$ for an optimal choice of $R_0$ with the models of \citet{jackson2017} and \citet{kolb1990}. In the optically-thin regime ($\delta R_{\rm{d}} < 0$) we do the comparison with $\dot{M}_{\rm{J}}$ (equation~\ref{eq:Mdot_J}), while in the optically-thick regime ($\delta R_{\rm{d}} > 0$) with $\dot{M}_{\rm{KR}}$  (equation~\ref{eq:Mdot_KR}). We now measure the absolute radius excess $\Delta R_\dd$ in the units of the photosphere pressure scale height $H_{P,\ph}$ defined as
\begin{equation}
    H_{P,\ph} \equiv \frac{P_\ph}{\rho_\ph} \frac{R_\dd^2}{G M_\dd}.
\end{equation}
For our main sequence donor we obtain $H_{P,\ph} = 1.81 \times 10^{-4}R_\dd = 1.72 \times 10^{-4}\,\Rsun$. In the upper panel, we see that $\dot{M}_\new$ has a smoother derivative than the standard prescription around $\Delta R_\dd/H_{P,\text{ph}} \approx 0$, because we do not have any artificial optically thin-thick transition in our model. In the lower panel, we show the ratios $\dot{M}_\text{new}/\dot{M}_\text{J}$ and $\dot{M}_\text{new}/\dot{M}_\text{KR}$  and we see that our model predicts roughly twice as large MT rate for small radius excess $\Delta R_{\rm{d}}/H_{P,\ph} \approx 0 $, but is similar to existing models for $\Delta R_{\rm{d}} /H_{P,\ph} \gtrsim 1$.

\subsubsection{Sun-like donor on the red giant branch}
\label{sec:comp_rgb}
In the right column of Fig.~\ref{fig:Mdot-DR-1M}, we show the results for our red giant donor with photosphere pressure scale height $H_{P,\ph} = 0.011 R_\dd = 0.96\,\Rsun$. We were unable to obtain solutions for $\Delta R_\dd/H_{P,\ph} \approx 0$ for $\Delta N_{H_P} < 1.25$. Since our analysis in Section~\ref{sec:sun-like-rgb_R0} indicated that the ideal choice of $\Delta N_{H_P}$ is between 0.5 and 1.0, our results for the red giant will be somewhat affected by the varying entropy in the surface layers. From our results we can draw similar conclusions as for the main sequence donor, specifically, that for $\Delta R_\dd/H_{P,\ph} \approx 0$ our model predicts roughly twice as large MT rate relative to \citet{kolb1990} and that our new and existing models are similar for  $\Delta R_\dd/H_{P,\ph} \gtrsim 2$.

\subsubsection{$30\,\Msun$ donor undergoing thermal time-scale mass transfer}
\label{sec:30-Msun_comp}

\begin{figure}
    \centering
        \includegraphics[width=\columnwidth]{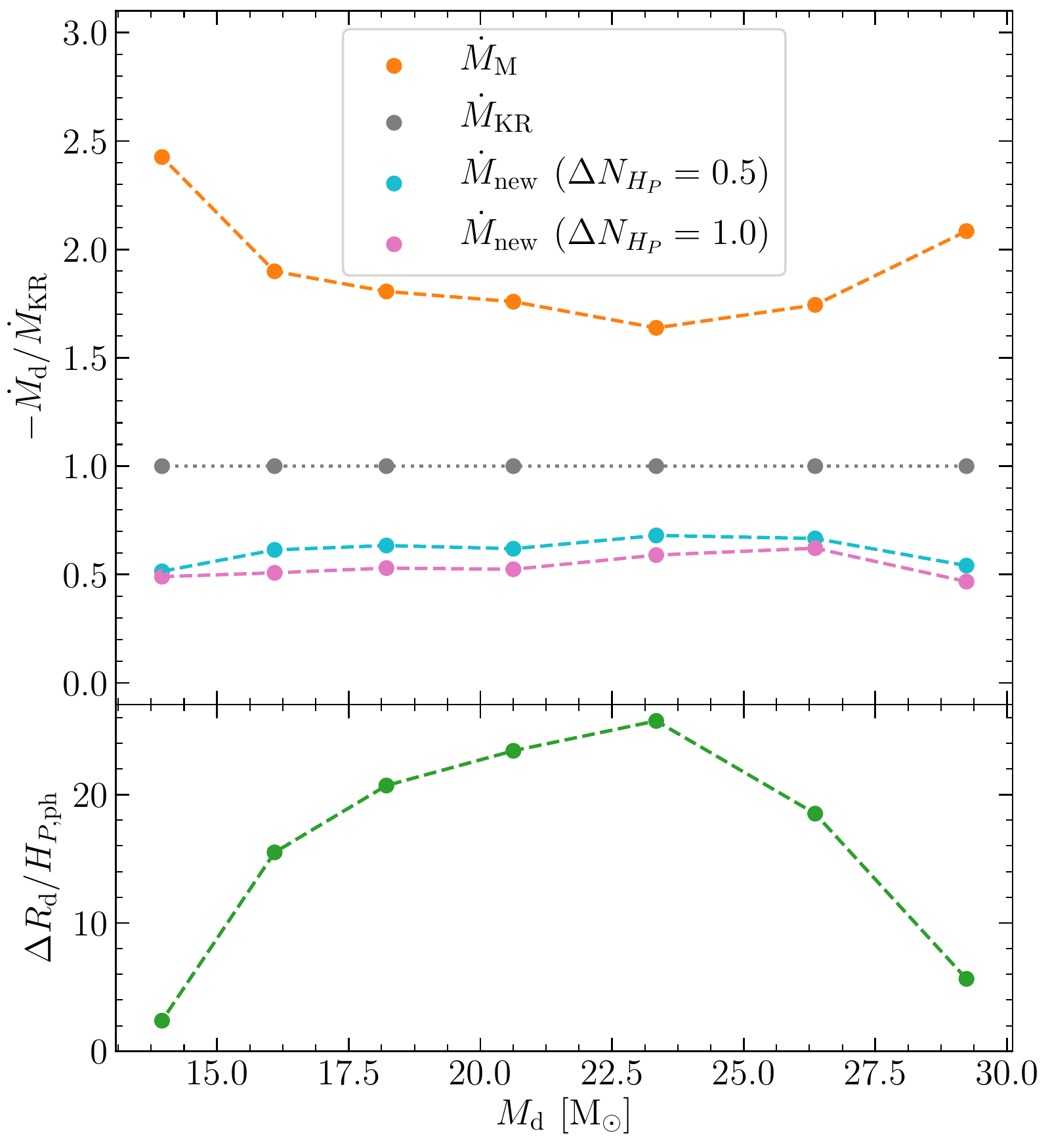}
    \caption{The comparison of different MT rate prescriptions relative to $\dot{M}_{\rm{KR}}$ for different evolutionary stages of the $30 \Msun$ low-metallicity donor undergoing thermal time-scale MT \citep{marchant2021}. The donor is evolved using the full MT prescription $\dot{M}_{\rm{M}}$ developed by \citet{marchant2021}. The MT rates $\dot{M}_{\rm{KR}}$ (equation~\ref{eq:Mdot_KR}) and $\dot{M}_{\rm{new}}$ (equation~\ref{eq:Mdot}) are computed a posteriori. The radius excess $\Delta R_{\rm{d}}$ in units of $H_{P,\rm{ph}}$ is shown in the lower panel.}
    \label{fig:Mdot_comp-m-KR-new}
\end{figure}

We evolve the $30\,\Msun$ donor using the files provided by \citet{marchant2021}\footnote{We used the latest Version 3 data set available at \url{https://doi.org/10.5281/zenodo.4574367} and we rerun some of the simulations using the same MESA version r15140 and MESA SDK version x86\_64-linux-20.12.1. However, we find small varying differences sometimes on the level of tens of per cent. We are not aware of the reason for this discrepancy, but we assume that it will not significantly impact our comparison.}, where MT is calculated using prescription $\dot{M}_\text{M}$. We investigate profiles of the  donor in several stages of its MT evolution and we compute $\dot{M}_{\rm{KR}}$ and $\dot{M}_{\rm{new}}$ for $\Delta N_{H_P}=0.5$ and $\Delta N_{H_P}=1.0$. In Fig.~\ref{fig:Mdot_comp-m-KR-new}, we show the comparison of different MT rate prescriptions and the degree of Roche lobe overflow measured with $\Delta R_\text{d}/H_{P,\text{ph}}$. We find that it roughly holds $\dot{M}_{\rm{M}} \approx 2.0 \dot{M}_{\rm{KR}}$ and $\dot{M}_{\rm{new}} \approx 0.55 \dot{M}_{\rm{KR}}$ during the whole evolution. This factor of 2 difference approximately equals to the maximum differences obtained for a range of $\Delta R_\text{d}$ for the $1\,\Msun$ donor, but the differences in these two situations are in the opposite direction. In all cases, the maximal differences occur for $\Delta R_\text{d}/H_{P,\text{ph}} \approx 0$, which is where the existing prescriptions stitch together the optically-thin and optically-thick regimes.

\subsubsection{Effects of our new model on the evolution of $30\,\Msun$ donor}
\label{sec:comp_evol}

\begin{figure}
    \centering
	\includegraphics[width=\columnwidth]{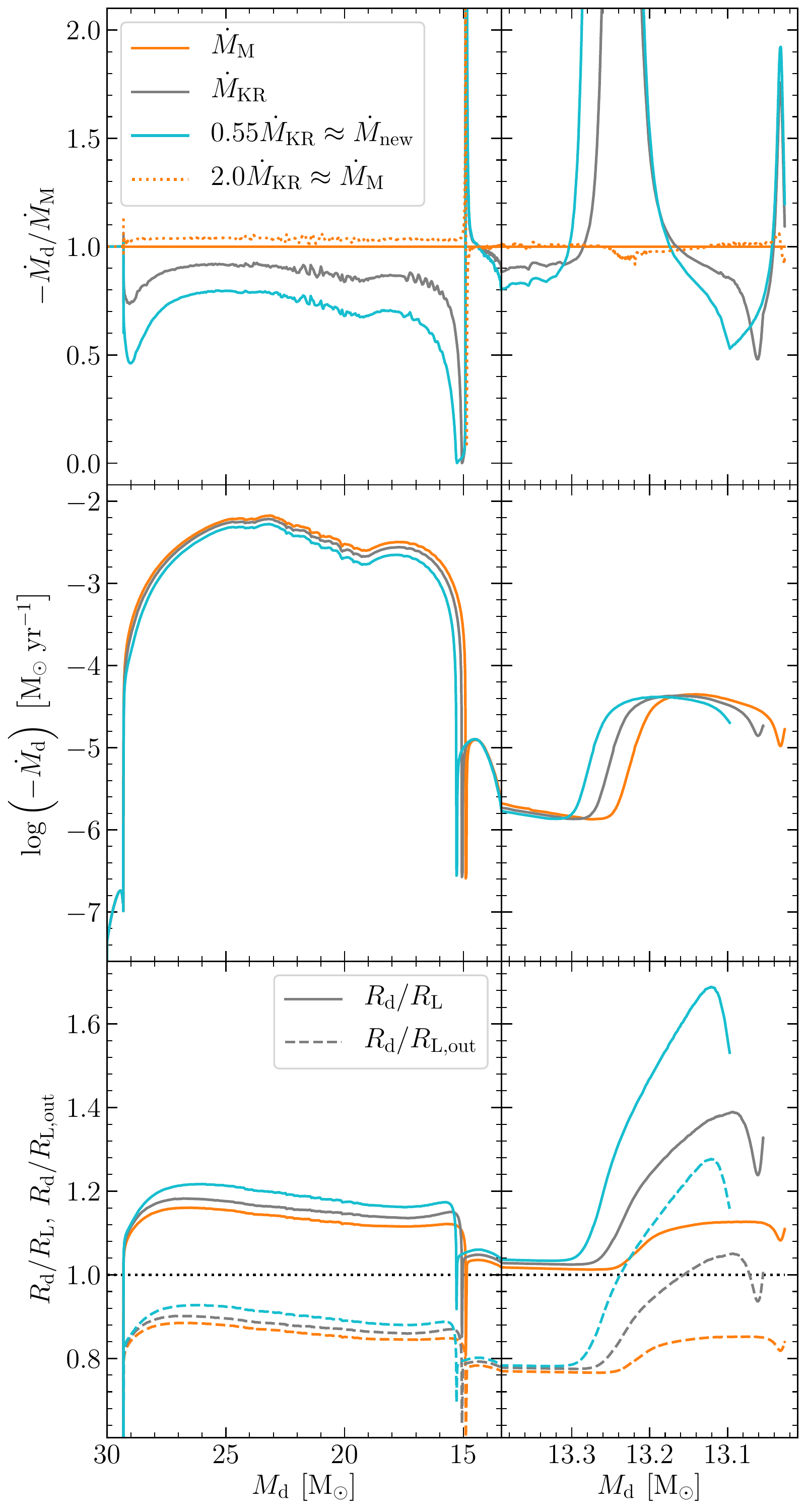}
    \caption{Evolution of the $30\,\Msun$ donor using different MT rate prescriptions based on \citet{marchant2021}. Top row shows the MT rate relative to the scheme of \citet{marchant2021}, middle row shows the absolute values of MT rates, and the bottom row shows the donor radius relative to the equivalent radii of L1 and L2 points. Left column shows almost the full evolution while the right column shows a detail of the second phase of MT.}
    \label{fig:Mdot-R-m-KR-new}
\end{figure}

Implementing self-consistently our model in 1D stellar evolution code like MESA is beyond the scope of this work. However, we can use the fact that for $30\,\Msun$ donor undergoing intensive thermal time-scale MT the value of $\dot{M}_\text{new}$ is consistently by a factor of two smaller than $\dot{M}_\text{KR}$, which is another factor of two smaller than $\dot{M}_\text{M}$, and we can learn about the effects of our new MT model in an approximate way. First, we evolve the $30\,\Msun$ donor using the modified \texttt{'Kolb'} MT prescription with a prefactor of $2.0$ to mimic $\dot{M}_{\rm{M}}$. Second, we use a prefactor of  $0.55$ to mimic $\dot{M}_{\rm{new}}$. In Fig.~\ref{fig:Mdot-R-m-KR-new}, we show the results of this experiment. Looking at the top row showing ratio with respect to $\dot{M}_\text{M}$, we see that MT rate of $2.0\dot{M}_\text{KR}$ closely resembles $\dot{M}_\text{M}$. This gives credibility to our approximate way of modifying the MT scheme and suggests that $0.55 \dot{M}_{\rm{KR}}$ represents the evolution of this specific donor if we self-consistently used our MT rate prescription $\dot{M}_{\rm{new}}$. 

In the middle row of Fig.~\ref{fig:Mdot-R-m-KR-new}, we show the absolute value of the MT rate and we see that the evolution tracks are qualitatively very similar and achieve similar MT rates. This is expected, because the donor's structure and binary properties determine the desired MT rate. However, there are significant differences in the evolutionary outcomes, which we investigate in the bottom row of Fig.~\ref{fig:Mdot-R-m-KR-new} by looking at the  position of stellar photosphere with respect to L1 and L2 points. We see that the model with $\dot{M}_{\rm{M}}$ does not predict any overflow of the  outer Lagrangian point, but that models  with $\dot{M}_{\rm{KR}}$ and an approximation for $\dot{M}_{\rm{new}}$ result in an overflow of L2 during the second MT phase. Model with approximated $\dot{M}_\text{new}$ predicts significantly higher L2 overflow than the model with $\dot{M}_\text{KR}$. This is expected, because our model requires higher $\delta R_\text{d}$ to achieve about the same $\dot{M}_\text{d}$.  L2 overflow is typically associated with instability leading to common-envelope evolution and thus our new MT rate scheme makes the MT less stable as compared to standard models \citep{kolb1990,marchant2021}, at least in this specific case.

\subsubsection{Ambiguity in evaluation of potentials}\label{sec:potential_ambiguity}

\begin{figure}
    \centering
	\includegraphics[width=\columnwidth]{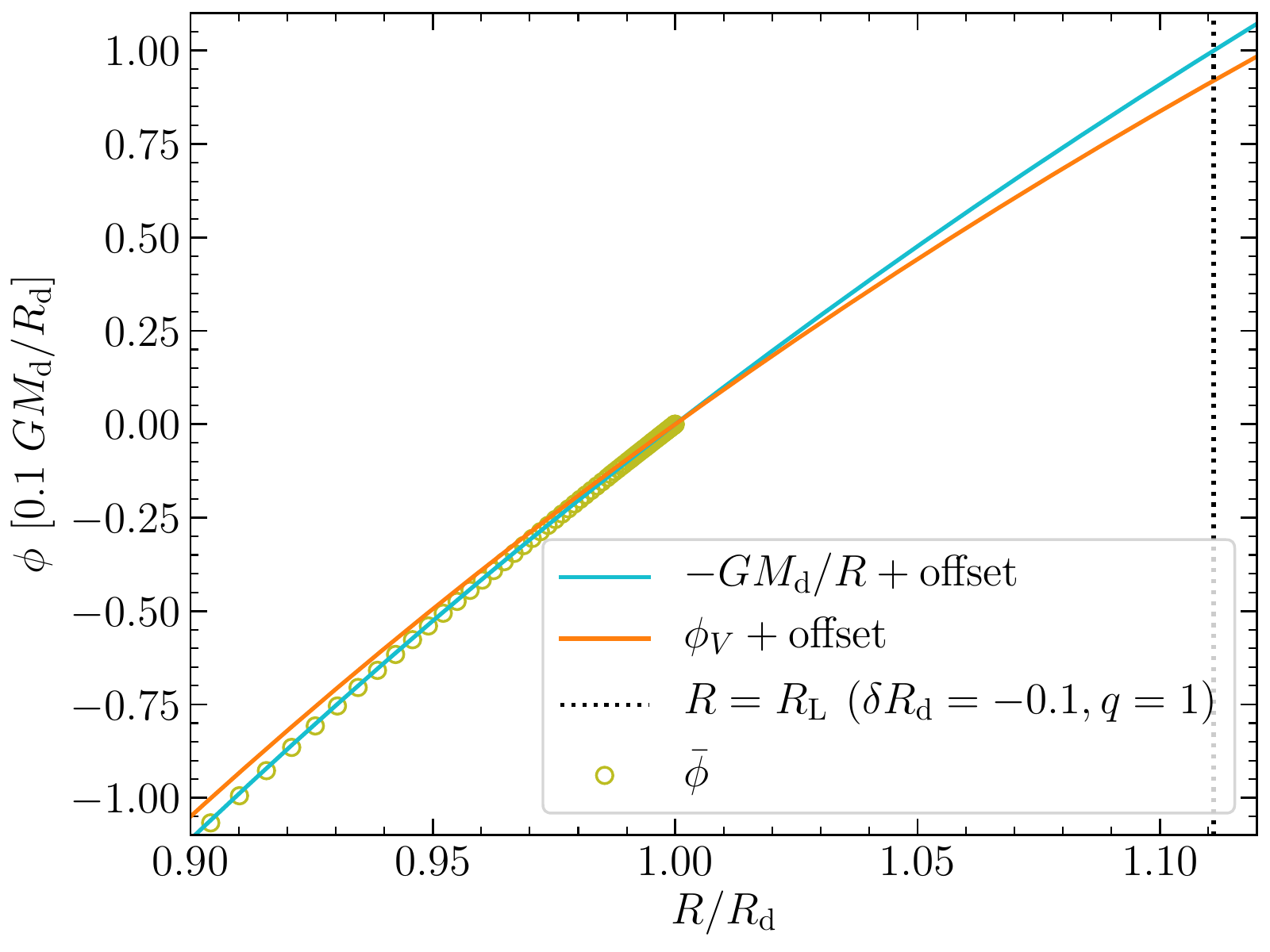}
    \caption{Different approximations for the potential near the outer layers of the $1\,\Msun$ donor on the main sequence for one specific case of $\delta R_\dd = -0.1$ and $q=1$. The potential $\phi_V$ is given by equation~(\ref{eq:phi_vol}) and the potential $\bar{\phi}$ by equation~(\ref{eq:phi_thick}). If needed, we add offsets to the potential to ensure $\phi(R_\dd) = 0$.}
    \label{fig:phi_1M}
\end{figure}

Finally, we address an ambiguity in connecting MT models with outer layers of hydrostatic stellar models through the potential. This ambiguity arises, because 1D hydrostatic models are typically evolved assuming that equipotentials of $\bar{\phi}$ are spheres. However, this is not the case for Roche potential, which is reflected in $\phi_V$. In Fig.~\ref{fig:phi_1M}, we show the potentials as a function of the radius $R$ for a Roche-lobe underfilling $1\,\Msun$ donor on the main sequence,  but the trends are the same for other types of donors. The more precise approximation of the Roche potential outside of the donor is the volume-equivalent potential $\phi_V$. The more precise approximation of the potential in the donor outer layers is the simple potential $\phi(R) \propto - G M_\dd/R$ because we evolved the donor as a single spherically symmetric non-rotating star. The potential $\phi_V$ is shallower than the potential $\phi \lp R\rp \propto - G M_\dd/R$, which implies that the donor would be more inflated if it were evolved in Roche geometry. Larger radius of the donor can further decrease the stability of MT.  

To check what is the effect of potential on the optically-thin rate $\dot{M}_{\rm{J}}$ (equation~\ref{eq:Mdot_J}), we calculate $\dot{M}_\text{J}$ for a simplified potential difference
\begin{equation}
    \phi_1 - \phi_\ph = -G \frac{M_\dd}{R_{\rm{L}}} + G \frac{M_\dd}{R_\dd} = - G \frac{M_\dd}{R_\dd} \delta R_\dd.
\end{equation}
In Fig.~\ref{fig:Mdot-DR-1M}, we show the resulting modification of the MT rate. We see that using the simplified version of $\dot{M}_{\rm{J}}$ makes the standard MT prescription smoother by lowering the jump in the derivative at $\Delta R_\text{d}/H_{P,\text{ph}} = 0$. This occurs because the simple approximation of the potential is closer to the hydrostatic potential $\bar{\phi}$ in the outer layers of the donor than $\phi_V$. 

The ambiguity would be removed if we evolved the donor in the Roche geometry in the 1D stellar evolution code so that $\phi_V$ would become the true potential. This was recently achieved by \citet{fabry2022} who found that there are two reasons why the radius of a tidally locked donor nearly  filling its Roche lobe increases compared to a single non-rotating star. First, the donor inflates by $\sim$5 per cent in radius simply due to co-rotation with the companion. Second, an additional  radius change of $\sim$1 per cent occurs due to tidal deformation from the companion.

\section{Summary and discussion}\label{sec:discussion}

In this work, we have developed a new model of MT in binary stars.  We argued that the Roche potential creates a de Laval-like nozzle around the L1 point and that gas flows primarily along the line connecting both stars and across equipotentials (Fig.~\ref{fig:bmt_scheme}). We formulated our equations by starting with 3D time-steady Euler equations and averaged them over the plane perpendicular to the gas motion. We assume that the gas is in hydrostatic equilibrium in the perpendicular plane. We obtained a set of equations describing 1D two-point boundary value problem starting at some point $R_0$ below the donor's surface and ending at L1 (equations~\ref{eq:hydro_gen}). The MT rate is the eigenvalue of the equations (equation~\ref{eq:Mdot}), similarly to how mass-loss rate is calculated in stellar winds. We obtained algebraic solutions for density and velocity profiles for isothermal (Sec.~\ref{sec:res_iso}, equation~(\ref{eq:Mdot_iso}), Fig.~\ref{fig:iso-v_rho_prof}) and ideal gas (Sec.~\ref{sec:res_id}, equation~(\ref{eq:Mdot_id}), Fig.~\ref{fig:id-v_rho_prof}). For realistic EOS from MESA, we obtained numerical solutions using relaxation (Sec.~\ref{sec:res_real}, equation~(\ref{eq:Mdot}), Fig.~\ref{fig:real_v_id-prof}). By carefully analyzing our solutions, we showed that for isothermal gas our model reduces to the existing optically-thin model of \citet{ritter1988} and \citet{jackson2017} (Sec.~\ref{sec:comp_iso}, equation~(\ref{eq:iso_Mdot_comp}), Fig.~\ref{fig:iso_comp}). For ideal gas and polytropic donor, we analytically showed that our MT rate differs from the existing optically-thick model of \citet{kolb1990} by a factor, which only weakly depends on adiabatic index $\Gamma$ (Sec.~\ref{sec:comp_ideal}, equation~(\ref{eq:id_Mdot_comp}), Fig.~\ref{fig:id_comp}). This factor peaks at 1.00 and has a minimum around 0.93 for the range of astrophysically-plausible values of $\Gamma$ (Fig.~\ref{fig:id_comp_lim_max}, equation~\ref{eq:id_Mdot_comp_lim_max}). 

For a realistic EOS, we applied our model to three realistic donors calculated in MESA (Sec.~\ref{sec:comp_real}). For $1\,\Msun$ donor on the main sequence or on the red giant branch our model predicts MT rates up to a factor of $2$ higher than in existing models (Fig.~\ref{fig:Mdot-DR-1M}). For an initial $30\,\Msun$ low-metallicity donor undergoing intensive thermal time-scale MT studied by \citet{marchant2021}, we found MT rates about a factor of $2$ smaller than in the model of \citet{kolb1990} and about a factor of $4$ smaller than in the model of \citet{marchant2021} (Fig.~\ref{fig:Mdot_comp-m-KR-new}). To estimate the effect of our new MT model on binary evolution, we calculated the evolution of $30\,\Msun$ donor of \citet{marchant2021} but with MT rate artificially lowered. We found that the donor overflows the outer Lagrange point of the binary, which is likely followed by a rapid orbital decay and common-envelope evolution. This contrasts with the original model, where the donor never overflowed L2 and recovered from the MT. It remains to be seen what our MT model predicts for a wider range of donors, but taking our results at face value suggests that any binary evolution feature or outcome should be robust against variations of donor's MT rate by a factor of about 2.

Our new model has several advantages with respect to existing MT models (Sec.~\ref{sec:review}). First, our model does not have the artificial division between optically-thick and optically-thin regimes, yet, it naturally provides results converging to existing models in the isothermal and polytropic limits.  Second, our new model is formulated as an eigenvalue problem so that all of donor's subsonically connected interior determines the MT rate. This is similar to how the mass-loss rate of stellar winds is calculated. Third, by reconstructing structure of the donor near L1, our model can capture phenomena that would not be predicted by 1D hydrostatic models. In particular, energy transfer in 1D stellar models occurs either by diffusion or convection, but in our model energy can be transferred by advection in the vicinity of L1.  Finally, we can add extra physics to our model by modifying the Euler equations~(\ref{eq:hydro}) and applying our averaging procedure. One possible extension is to start with equations of magnetohydrodynamics to illuminate the interaction between surface magnetic fields and MT, which would be of interest for binaries with low-mass donors such as CVs and X-ray binaries. 

However, a more important addition to our model would be some treatment of radiation transport. We know that the optically-thin flow occurs for optical depths $\tau \ll 1$, but the optically-thick flow, $\tau \gg 1$, can be further divided in two regimes based on the flow velocity. If $\tau \gg c/v$, radiation is effectively trapped and moves together with the gas. This regime is traditionally called optically-thick MT, but really the key assumption is the adiabaticity rather than optical depth. If $1 \ll \tau \ll c/v$, the radiation diffuses faster than the gas flows and the adiabaticity is violated. This intermediate regime occurs in stellar interiors as well as other situations \citep[e.g.,][]{krumholz07,piro20,calderon21}, but is currently  neglected in the theory of MT. To illustrate this point, we can estimate the critical $\tau \sim c/v \sim 4\times 10^4$ from the sound speed at a typical $6000$\,K stellar photosphere. We plan to include the treatment of radiation transport in future work.

Some of the assumptions built into our model are similar to the ones in existing models, but we apply them differently. One example is the assumption that the flow is adiabatic along the streamlines, where existing optically-thick models assume streamlines aligned with equipotentials, but our model does not. Another example is our assumption of polytropic structure perpendicular to the orbital plane, which facilitates closed form of equations~(\ref{eq:hydro_gen}). This is similar to the polytropic assumption in the model of \citet{kolb1990}. Similarly to \citet{pavlovskii2015}, who replaced polytropic assumption along the streamline with a realistic EOS, we could use realistic EOS for the perpendicular hydrostatic structure. The most convenient form to do this would be with a pre-calculated table on a grid of $\rho$ and $T$. We leave this potential improvement for future work. To summarize, our new MT model is built on a set of assumptions, which are mostly different but also partially overlap with assumptions of existing models. As a result, our effort can be interpreted as a probe of one of the systematic uncertainties in binary evolution calculations \citep[e.g.,][]{mandel2022}.

One open issue is how feasible is to include our new model in stellar evolution codes such as MESA. Here, the existing models have a clear advantage, because they provide relatively simple prescriptions, which operate on calculated thermodynamic profiles and do not require solving additional equations. Nonetheless, our equations could be implemented within a code like MESA, because the required routines for solving two-point boundary value problem and calling the EOS are already included. Since our model does not involve any nuclear reactions and requires fewer grid points than the full stellar model, the computational penalty would be at worst a factor of about 2. However, a simpler way would be to pre-calculate MT rates for a grid of stellar envelope parameters such as $\rho$, $T$, and metallicity, and load the results in MESA as a table.

Finally, let us briefly mention two possible applications of our model. First, existing models assume that the mass is lost in approximately spherically symmetric fashion from the donor and is then transferred to L1 (Fig.~\ref{fig:bmt_scheme_KR}), however, our new model assumes that the mass is lost from one side of the donor facing the accretor (Fig.~\ref{fig:bmt_scheme}). If the donor is not co-rotating with the orbit, the MT will still happen but the position of a point similar to L1 will be different \citep{sepinsky2007}. The MT from a rotating donor leads to the loss of donor's rotational angular momentum, which can affect the spin vector, especially during rapid MT phases \citep{matese1983,stegmann2021}. Second, intense irradiation of the donor from the accretor can provide sufficient external radiation pressure so that the L1 point can disappear entirely  and the mass is lost through L2 with consequences for binary stability \citep{phillips2002}. Another interesting effect of radiation pressure would occur in massive donors, which  feature locally super-Eddington luminosities due an iron opacity bump close to their surface \citep[e.g.,][]{jiang15}. The local decrease of gravity near L1 could cause an enhanced MT rate in these donors. We plan to investigate all of these issues once we include radiation transport in our MT model.

\section*{Acknowledgements}

This work has been supported by Horizon 2020 ERC Starting Grant 'Cat-In-hAT' (grant agreement no. 803158). Most of the algebraic calculations and visualizations in this work were performed with Mathematica 12.2 \citep{mathematica12.2}, Matplotlib \citep{hunter2007}, and NumPy \citep{harris2020}.


\section*{Data Availability}

The data underlying this article will be shared on reasonable request to the corresponding author.


\bibliographystyle{mnras}
\bibliography{bibliography}



\appendix

\section{Derivation of 1D hydrodynamic equations}\label{sec:hydro_gen}
Here, we show additional steps in the derivation of 1D hydrodynamic equations~(\ref{eq:hydro_gen}) from the general 3D Euler equations~(\ref{eq:hydro}) using the assumptions (i)-(v) stated in Section~\ref{sec:hydro}. Along the way, we derive the expressions for the effective density and pressure cross-sections (equations~\ref{eq:Q_rho} and \ref{eq:Q_frac}) defined in equation~(\ref{eq:Q_def}). By using the stationarity assumption (i) and equation~(\ref{eq:eps_tot}), we get
\begin{subequations}
	\label{eq:hydro_i}
	\begin{align}
	\boldsymbol{\nabla} \boldsymbol{\cdot} \left(\rho \boldsymbol{v}\right) &= 0 , \label{eq:hydro_i_mass} \\
	\boldsymbol{\nabla} \boldsymbol{\cdot}\left(\rho \boldsymbol{v} \otimes \boldsymbol{v} + P \boldsymbol{\mathsf{I}} \right) &= -\rho \boldsymbol{\nabla} \phi_{\rm R}, \label{eq:hydro_i_mom} \\
    \boldsymbol{\nabla} \boldsymbol{\cdot} \left[ \left(\rho \epsilon + \frac{1}{2} \rho \boldsymbol{v}^2 + \rho \phi_{\rm R} + P \right) \boldsymbol{v}\right] &= 0. \label{eq:hydro_i_en}
	\end{align}
\end{subequations}
Now, we focus on the gas flow along the $x$ axis. We use a weaker definition of the gas cross-section $Q$ in the $yz$ plane, $\rho \left(\partial Q\right) \approx 0$, because we do not have the polytropic assumption in the $yz$ plane yet. Hence, the gas does not flow through the boundary $\partial Q$. Thus, the mass-flow rate through the cross-section $Q\left(x\right)$ is conserved along the $x$ axis and the mass equation (\ref{eq:hydro_i_mass}) reduces to
\begin{equation}
    \frac{\dd}{\dd x} \int_{Q(x)} \rho(x,y,z) v_x(x,y,z) \dd Q = 0.
\end{equation}
Using assumption (ii) and the definition of the effective density cross-section in equation~(\ref{eq:Qrho}) we get
\begin{equation}
    \frac{\dd}{\dd x} \left(v_x(x) \rho(x,0,0) Q_\rho(x) \right) = \frac{\dd}{\dd x} (v_x \rho_{\rm c} Q_\rho) = 0,
    \label{eq:hydro_1D_mass}
\end{equation}
where we denote the value of density on the $x$-axis by $\rho(x,0,0) \equiv \rho_{\rm c}(x) = \rho_\text{c}$. We see that the mass equation~(\ref{eq:hydro_1D_mass}) is equivalent to the mass equation~(\ref{eq:hydro_gen_mass}).

The $x$ component of the momentum equation~(\ref{eq:hydro_i_mom}) reads
\begin{equation}
    \frac{\partial}{\partial x} (\rho v_x^2 + P) + \frac{\partial}{\partial y} (\rho v_x v_y) + \frac{\partial}{\partial z} (\rho v_x v_z) = -\rho \frac{\partial \phi_{\rm R}}{\partial x}.
\end{equation}
Using assumption (iii), we can get rid of two terms in this equation to get
\begin{equation}
    \frac{\partial}{\partial x} (\rho v_x^2 + P) = -\rho \frac{\partial \phi_{\rm R}}{\partial x}.
\end{equation}
Because at the boundary $\partial Q$ we have $\rho(\partial Q) \approx 0$, it also has to hold that $P(\partial Q) \approx 0$. Therefore, by integrating the equation over the cross-section $Q\left(x\right)$ and using assumption (iv) we arrive at
\begin{equation}
        \frac{\dd}{\dd x} \int_{Q(x)} \left(\rho v_x^2 + P \right) \dd Q = -\frac{\dd \phi_{\rm R}}{\dd x} \int_{Q(x)} \rho \dd Q.
\end{equation}
Using assumption (ii) and the definitions of the effective density and pressure cross-sections from equation~(\ref{eq:Q_def}) gives us
\begin{equation}
        \frac{\dd}{\dd x}  (v_x^2 \rho_{\rm c} Q_\rho) + \frac{\dd}{\dd x} (P_{\rm c} Q_P) = - \rho_{\rm c} Q_\rho \frac{\dd \phi_{\rm R}}{\dd x},
\end{equation}
where we denote the value of pressure on the $x$ axis by $P(x,0,0) \equiv P_{\rm c}(x) = P_{\rm c}$. Dividing the equation by $\rho_{\rm c} Q_\rho$ and using the mass equation~(\ref{eq:hydro_1D_mass}) finally leads to the momentum equation~(\ref{eq:hydro_gen_mom}).

Applying assumption (iii) on the energy equation~(\ref{eq:hydro_i_en}) gives
\begin{equation}
    \frac{\partial}{\partial x} \left[ \left(\rho \epsilon + \frac{1}{2} \rho v_x^2 + \rho \phi_{\rm R} +P\right)v_x\right]=0.
\end{equation}
Using assumption (ii), integrating over the cross-section $Q(x)$, and keeping in mind that $\rho(\partial Q) \approx 0, P(\partial Q) \approx 0$, yields
\begin{equation}
    \frac{\dd}{\dd x} \left[ v_x \int_{Q(x)} \left(\rho\epsilon + \frac{1}{2} \rho v_x^2 + \rho \phi_{\rm R} + P \right) \dd Q \right] = 0.
\end{equation}
Using the definitions of the effective density and pressure cross-sections in equation~(\ref{eq:Q_def}), the mass equation (\ref{eq:hydro_1D_mass}), and dividing the equation by $v_x \rho_{\rm c} Q_\rho$ gives
\begin{equation}
    \frac{\dd}{\dd x} \left[ \frac{1}{2}v_x^2 + \frac{P_{\rm c} Q_P}{\rho_{\rm c} Q_\rho} + \frac{1}{\rho_{\rm c} Q_\rho} \int_{Q(x)} (\rho \epsilon + \rho \phi_{\rm R}) \dd Q \right] = 0.
    \label{eq:hydro_1D_en_undone}
\end{equation}

Now, we want to evaluate the term including the internal energy per mass unit $\epsilon$. This term accounts for change in internal energy as gas passes along the $x$ axis, therefore we want to express this term as internal energy on the $x$ axis $\epsilon(x,0,0) \equiv \epsilon_{\rm c}(x) = \epsilon_{\rm c}$ times some dimensionless factor. Assumption (v) also implies $\epsilon = P / [(\Gamma-1)\rho]$ in the $yz$ plane. Thus, we can write
\begin{equation}
    \begin{aligned}
        \frac{1}{\rho_{\rm c} Q_\rho} \int_{Q(x)} \rho \epsilon \dd Q &= \frac{1}{\rho_{\rm c} Q_\rho} \frac{1}{\Gamma-1} \int_{Q(x)} P \dd Q \\
        &= \frac{1}{\rho_{\rm c} Q_\rho} \frac{1}{\Gamma-1} P_{\rm c} Q_P = \epsilon_{\rm c} \frac{Q_P}{Q_\rho},
    \end{aligned}
    \label{eq:hydro_1D_en_epsilon}
\end{equation}
where of course the fifth assumption (v) applies to the whole $yz$ plane, hence it also holds on the $x$ axis, $\epsilon_{\rm{c}} = P_{\rm{c}} / [(\Gamma-1)\rho_{\rm{c}}]$. We can also express this term as the isothermal sound speed squared on the $x$ axis, $c_T^2(x,0,0) \equiv c_{T,\rm c}^2(x) = c_{T,\rm c}^2$, times some other dimensionless factor, or alternatively the fraction $P_{\rm c} / \rho_{\rm c}$ times another dimensionless factor. Our aim is to keep the polytropic approximation only in the $yz$ plane and to allow for a general EOS along the $x$ axis. Thus, we keep the internal energy per unit mass $\epsilon_{\rm c}$ in this term. 

Using assumptions (iii), (iv), and (v), we can express the Roche potential in the following form
\begin{equation}
    \phi_{\rm R} - \phi_{\rm R}^x = \frac{K \Gamma}{\Gamma-1} \left(\rho_{\rm c}^{\Gamma-1}-\rho^{\Gamma-1}\right) = \frac{ \Gamma}{\Gamma-1} \left(c_{T,\rm c}^2-c_T^2\right).
    \label{eq:pot_yz}
\end{equation}
Therefore, we can express the potential term in equation~(\ref{eq:hydro_1D_en_undone}) as
\begin{equation}
    \begin{aligned}
        &\frac{1}{\rho_{\rm c} Q_\rho} \int_{Q(x)} \rho \phi_{\rm R} \dd Q \\
        & \quad = \frac{1}{\rho_{\rm c} Q_\rho} \int_{Q(x)} \left(\phi_{\rm R}^x\rho + \frac{K \Gamma}{\Gamma-1} \rho_{\rm c}^{\Gamma-1} \rho - \frac{K \Gamma}{\Gamma-1} \rho^\Gamma \right)\dd Q \\
        & \quad = \phi_{\rm R}^x(x) + \frac{\Gamma}{\Gamma-1} \frac{P_{\rm c}}{\rho_{\rm c}} - \frac{1}{\rho_{\rm c} Q_\rho}\frac{\Gamma}{\Gamma-1} P_{\rm c} Q_P \\
        & \quad = \phi_{\rm R}^x(x) + c_{T,\rm c}^2 \frac{\Gamma}{\Gamma-1} \left(1-\frac{Q_P}{Q_\rho}\right),
    \end{aligned}
    \label{eq:hydro_1D_en_potential}
\end{equation}
where we choose to evaluate the potential term as function of the potential on the $x$ axis $\phi_{\rm R}^x$ times some dimensionless factor and the isothermal sound speed squared on the $x$ axis $c_{T,\rm c}^2$ times another dimensionless factor. 

By combining the expression for the internal energy term (equation~\ref{eq:hydro_1D_en_epsilon}) and the potential term (equation~\ref{eq:hydro_1D_en_potential}) in the energy equation~(\ref{eq:hydro_1D_en_undone}), we get
\begin{equation}
    \begin{aligned}
    & v_x\frac{\dd v_x}{\dd x} + \frac{\dd}{\dd x} \left(\frac{1}{\rho_{\rm c} Q_\rho} P_{\rm c} Q_P\right) + \frac{\dd}{\dd x} \left(\epsilon_{\rm c} \frac{Q_P}{Q_\rho}\right) \\
    & \quad = -\frac{\dd}{\dd x} \left[\phi_{\rm R}^x + c_{T,\rm c}^2 \frac{\Gamma}{\Gamma-1} \left(1-\frac{Q_P}{Q_\rho}\right)\right].
    \end{aligned}
    \label{eq:hydro_1D_en_undone2}
\end{equation}
Subtracting the momentum equation~(\ref{eq:hydro_gen_mom}) from this equation leaves us with the energy equation in the form
\begin{equation}
    \frac{\dd}{\dd x} \left(\epsilon_{\rm c} \frac{Q_P}{Q_\rho}\right) - \frac{P_{\rm c} Q_P}{(\rho_{\rm c} Q_\rho)^2} \frac{\dd}{\dd x} (\rho_{\rm c} Q_\rho) = -\frac{\dd}{\dd x} \left[ c_{T,\rm c}^2 \frac{\Gamma}{\Gamma-1} \left(1-\frac{Q_P}{Q_\rho}\right)\right].
    \label{eq:hydro_1D_en}
\end{equation}

We use assumption (iv) in the form $\phi_{\rm R} = \phi_{\rm R}^x(x) + B y^2/2 + C z^2/2$ and equation~(\ref{eq:pot_yz}) to arrive at
\begin{equation}
    \rho = \left(1-\frac{\Gamma-1}{2 \Gamma}\frac{B y^2}{K \rho_{\rm c}^{\Gamma-1}}-\frac{\Gamma-1}{2 \Gamma}\frac{C z^2}{K \rho_{\rm c}^{\Gamma-1}} \right)^{\frac{1}{\Gamma-1}} \rho_{\rm c}.
\end{equation}
We can define the following parameters
\begin{subequations}
    \label{eq:yz_max}
    \begin{align}
        y_{\rm{max}} &= \sqrt{\frac{2 \Gamma K \rho_{\rm c}^{\Gamma-1}}{(\Gamma-1)B}} = \sqrt{\frac{2 \Gamma}{(\Gamma-1)B}}c_{T,\rm c}, \\
        z_{\rm{max}} &= \sqrt{\frac{2 \Gamma K \rho_{\rm c}^{\Gamma-1}}{(\Gamma-1)C}} = \sqrt{\frac{2 \Gamma}{(\Gamma-1)C}}c_{T,\rm c}, 
    \end{align}
\end{subequations}
to obtain
\begin{equation}
    \rho \left(x,y,z\right) = \left[1 - \left(\frac{y}{y_{\rm{max}}}\right)^2 - \left(\frac{z}{z_{\rm{max}}}\right)^2\right]^{\frac{1}{\Gamma-1}} \rho_{\rm c} \left(x\right).
\end{equation}
By defining new variables $r$ and $\varphi$ as $y(r, \varphi) = y_{\rm{max}} r \cos \varphi$ and $z(r, \varphi) = z_{\rm{max}} r \sin \varphi$, the expression for density further simplifies to
\begin{equation}
    \rho(x,r, \varphi) = (1 - r^2)^{\frac{1}{\Gamma-1}} \rho_{\rm c} (x).
    \label{eq:rho_r}
\end{equation}

Now, we are ready to compute the effective cross-sections $Q_\rho, Q_P$  defined by the equation~(\ref{eq:Q_def}),
\begin{equation}
    \begin{aligned}
        &\rho_{\rm c} Q_\rho \equiv \int_{Q(x)} \rho \dd Q = \int_0^{1}\int_0^{2\pi} \rho y_{\rm{max}} z_{\rm{max}} r \dd\varphi \dd r \\
        & \quad = 2\pi y_{\rm{max}} z_{\rm{max}} \rho_{\rm c} \int_0^1 (1-r^2)^{\frac{1}{\Gamma-1}} r \dd r = \frac{\Gamma-1}{\Gamma}\pi y_{\rm{max}} z_{\rm{max}} \rho_{\rm c},
    \end{aligned}
\end{equation}
where we used equation~(\ref{eq:rho_r}). Using equation~(\ref{eq:yz_max}) we get the density cross-section
\begin{equation}
    Q_\rho = \frac{\Gamma-1}{\Gamma}\pi y_{\rm{max}} z_{\rm{max}} = \frac{2\pi}{\sqrt{BC}} c_{T,\rm c}^2,
    \label{eq:Q_rho_1D}
\end{equation}
where we choose to express the cross-section as function of the isothermal sound speed squared on the $x$ axis $c_{T,\rm c}^2$. We see that we recover the equation (\ref{eq:Q_rho}).

From equation~(\ref{eq:rho_r}) follows that
\begin{equation}
    P (x, r, \varphi) = (1 - r^2)^{\frac{\Gamma}{\Gamma-1}} P_{\rm c}(x).
    \label{eq:P_r}
\end{equation}
Thus, we can write the pressure cross-section as
\begin{equation}
    \begin{aligned}
        & \rho_{\rm c} Q_P \equiv \int_{Q(x)} P \dd Q = \int_0^{1}\int_0^{2\pi} P y_{\rm{max}} z_{\rm{max}} r \dd\varphi \dd r  \\
        & \quad = 2\pi y_{\rm{max}} z_{\rm{max}} P_{\rm c} \int_0^1 (1-r^2)^{\frac{\Gamma}{\Gamma-1}} r \dd r = \frac{\Gamma-1}{2\Gamma-1}\pi y_{\rm{max}} z_{\rm{max}} P_{\rm c},
    \end{aligned}
\end{equation}
where we used the equation (\ref{eq:P_r}). Using equation~(\ref{eq:yz_max}) we get
\begin{equation}
    Q_P = \frac{\Gamma-1}{2\Gamma-1}\pi y_{\rm{max}} z_{\rm{max}} = \frac{\Gamma}{2\Gamma-1}\frac{2\pi}{\sqrt{BC}} c_{T,\rm c}^2,
\end{equation}
which together with equation~(\ref{eq:Q_rho_1D}) leads to the equation~(\ref{eq:Q_frac}). Using this expression for the cross-section fraction also gives us
\begin{equation}
    \frac{\Gamma}{\Gamma-1} \left(1-\frac{Q_P}{Q_\rho}\right) = \frac{\Gamma}{2\Gamma-1} = \frac{Q_P}{Q_\rho},
\end{equation}
which means that the energy equation (\ref{eq:hydro_1D_en}) and the energy equation (\ref{eq:hydro_gen_en}) are identical.

\section{Derivation of the matrix form of 1D hydrodynamic equations}\label{sec:hydro_gen_mat}

In this section, we derive the matrix form of 1D hydrodynamic equations~(\ref{eq:hydro_gen}). We assume a general EOS along the $x$ axis, or equivalently, we assume the existence of functions $c_T = c_T (\rho, T)$, $P = P(\rho, T)$, $\epsilon = \epsilon (\rho, T)$, $\Gamma = \Gamma(\rho, T)$. Equation~(\ref{eq:hydro_gen_mat}) can be expressed as
\begin{equation}
    \renewcommand*{\arraystretch}{1.5}
    \begin{pmatrix}
        m_{11} & m_{12} & m_{13} \\
        m_{21} & m_{22} & m_{23} \\
        m_{31} & m_{32} & m_{33} 
    \end{pmatrix}
    \begin{pmatrix}
        \frac{\dd \ln v}{\dd x} \\
        \frac{\dd \ln \rho}{\dd x} \\
        \frac{\dd \ln T}{\dd x}
    \end{pmatrix}
    =
    \begin{pmatrix}
        0 \\
        -\frac{\dd \phi_{\rm R}}{\dd x} \\
        0
    \end{pmatrix}
    ,
\end{equation}
where $m_{ij}$, $i=1,2,3$, $j=1,2,3$, are the elements of matrix $\boldsymbol{\mathsf{M}}$. We denote the cross-section fraction $Q_P/Q_\rho$ determined by equation~(\ref{eq:Q_frac}) and its derivative by
\begin{equation}
    g \equiv \frac{Q_P}{Q_\rho} = \frac{\Gamma}{2\Gamma-1}, \quad g_\Gamma \equiv \frac{\dd g}{\dd \ln \Gamma} = - \frac{\Gamma}{(2\Gamma-1)^2}.
    \label{eq:g_def}
\end{equation}
The $\Gamma$ factor with its derivatives are
\begin{equation}
    \Gamma \equiv \left.\frac{\partial \ln P}{\partial \ln \rho} \right\vert_S, \quad \Gamma_\rho \equiv \left.\frac{\partial \ln \Gamma}{\partial \ln \rho} \right\vert_T, \quad \Gamma_T \equiv \left.\frac{\partial \ln \Gamma}{\partial \ln T} \right\vert_\rho.
    \label{eq:Gamma_def}
\end{equation}
Other useful thermodynamic quantities are defined as
\begin{equation}
    \begin{aligned}
        \chi_\rho &\equiv \left.\frac{\partial \ln P}{\partial \ln \rho}\right\vert_T, \quad \chi_T \equiv \left.\frac{\partial \ln P}{\partial \ln T}\right\vert_\rho, \quad \chi_{\rho \rho} \equiv \left.\frac{\partial \ln \chi_\rho}{\partial \ln \rho}\right\vert_T, \\
        \chi_{\rho T} &\equiv \left.\frac{\partial \ln \chi_\rho}{\partial \ln T}\right\vert_\rho, \quad \psi_\rho \equiv \left.\frac{\partial \ln \epsilon}{\partial \ln \rho}\right\vert_T, \quad \psi_T \equiv \left.\frac{\partial \ln \epsilon}{\partial \ln T}\right\vert_\rho.
    \end{aligned}
    \label{eq:thermo_def}
\end{equation}
The definition of the isothermal sound speed is
\begin{equation}
    c_T^2 \equiv \left.\frac{\partial P}{\partial \rho}\right\vert_T = \chi_\rho \frac{P}{\rho}.
    \label{eq:c_T_def}
\end{equation}
Using the expressions for the effective density and pressure cross-sections (equations~\ref{eq:Q_rho} and \ref{eq:Q_frac}), we can rewrite our set of 1D hydrodynamic equations into the form
\begin{subequations}
	\begin{align}
	\frac{\dd}{\dd x} (\chi_\rho v P) &= 0, \label{eq:hydro_mat_mass} \\
	v\frac{\dd v}{\dd x} + \frac{1}{\chi_\rho P} \frac{\dd}{\dd x} \left(g \chi_\rho \frac{P^2}{\rho} \right) &= -\frac{\dd\phi_{\rm R}}{\dd x}, \label{eq:hydro_mat_mom} \\
	\frac{\dd}{\dd x} (g \epsilon) - \frac{g}{\chi_\rho \rho} \frac{\dd}{\dd x} ( \chi_\rho P) + \frac{\dd}{\dd x} \left(g \chi_\rho \frac{P}{\rho} \right) &= 0. \label{eq:hydro_mat_en}
	\end{align}
	\label{eq:hydro_mat}
\end{subequations}

The mass equation (\ref{eq:hydro_mat_mass}) can be further expressed as
\begin{equation}
    \begin{aligned}
        &\frac{\dd \ln \chi_\rho (\rho,T)}{\dd x} + \frac{\dd \ln v}{\dd x} + \frac{\dd \ln P(\rho,T)}{\dd x} \\
        & \quad = \frac{\dd \ln v}{\dd x} + (\chi_\rho + \chi_{\rho \rho}) \frac{\dd \ln \rho}{\dd x} + (\chi_T + \chi_{\rho T}) \frac{\dd \ln T}{\dd x} = 0.
        \label{eq:hydro_mat_mass_final}
    \end{aligned}
\end{equation}
Thus, we have the first row of matrix $\boldsymbol{\mathsf{M}}$ as
\beq
    m_{11} = 1, \quad m_{12} = \chi_\rho + \chi_{\rho \rho}, \quad m_{13} = \chi_T + \chi_{\rho T}.
\eeq

The second term in the momentum equation~(\ref{eq:hydro_mat_mom}) can be written as
\beq
    \begin{aligned}
        &\frac{1}{\chi_\rho P} \frac{\dd}{\dd x} \lp g \chi_\rho \frac{P^2}{\rho}\rp = g_\Gamma \frac{P}{\rho} \frac{\dd \ln \Gamma}{\dd x} + g \frac{P}{\rho} \frac{\dd \ln \chi_\rho}{\dd x} + 2 g \frac{P}{\rho} \frac{\dd \ln P}{\dd x}  \\
        & \quad - g \frac{P}{\rho} \frac{\dd \ln \rho}{\dd x} = [g_\Gamma \Gamma_\rho + ( -1 + 2 \chi_\rho + \chi_{\rho \rho}) g ] \frac{P}{\rho} \frac{\dd \ln \rho}{\dd x}  \\
        & \quad + [g_\Gamma \Gamma_T + ( 2 \chi_T + \chi_{\rho T}) g ] \frac{P}{\rho} \frac{\dd \ln T}{\dd x}.
    \end{aligned}
    \label{eq:hydro_mat_mom_undone}
\eeq
Subtracting $g P/\rho$ times the mass equation~(\ref{eq:hydro_mat_mass_final}) from the momentum equation~(\ref{eq:hydro_mat_mom}) then leads to
\beq
    \begin{aligned}
        &\lp v^2 - g \frac{P}{\rho}\rp \frac{\dd \ln v}{\dd x} + [g_\Gamma \Gamma_\rho + ( -1 + \chi_\rho) g] \frac{P}{\rho} \frac{\dd \ln \rho}{\dd x} \\
        & \quad + ( g_\Gamma \Gamma_T + g \chi_T ) \frac{P}{\rho} \frac{\dd \ln T}{\dd x} = - \frac{\dd \phi_{\rm R}}{\dd x},
    \end{aligned}
    \label{eq:hydro_mat_mom_final}
\eeq
which gives us the second row of matrix $\boldsymbol{\mathsf{M}}$ as
\beq
    \begin{aligned}
        m_{21} &= v^2 - g \frac{P}{\rho}, \quad m_{22} = [g_\Gamma \Gamma_\rho + ( -1 + \chi_\rho ) g]\frac{P}{\rho},\\
        m_{23} &= ( g_\Gamma \Gamma_T + g \chi_T ) \frac{P}{\rho}.
    \end{aligned}
\eeq

For the first term in the energy equation~(\ref{eq:hydro_mat_en}) we have
\beq
    \begin{aligned}
        &\frac{\dd}{\dd x} ( g \epsilon ) = g_\Gamma \epsilon \frac{\dd \ln \Gamma}{\dd x} + g \epsilon \frac{\dd \ln \epsilon}{\dd x} \\
        & \quad = ( g_\Gamma \Gamma_\rho + g \psi_\rho ) \epsilon \frac{\dd \ln \rho }{\dd x} + ( g_\Gamma \Gamma_T + g \psi_T ) \epsilon \frac{\dd \ln T }{\dd x},
    \end{aligned}
\eeq
for the second term
\begin{equation}
    \begin{aligned}
        &- \frac{g}{\chi_\rho \rho} \frac{\dd}{\dd x} ( \chi_\rho P) = -g \frac{P}{\rho} \frac{\dd \ln \chi_\rho}{\dd x} - g \frac{P}{\rho} \frac{\dd \ln P}{\dd x} \\
        & = - ( \chi_\rho + \chi_{\rho \rho} ) g \frac{P}{\rho} \frac{\dd \ln \rho}{\dd x} - ( \chi_T + \chi_{\rho T} ) g \frac{P}{\rho} \frac{\dd \ln T}{\dd x},
    \end{aligned}
\end{equation}
and for the third term
\beq
    \begin{aligned}
        &\frac{\dd }{\dd x} \lp g \chi_\rho \frac{P}{\rho}\rp = g_\Gamma \chi_\rho \frac{P}{\rho} \frac{\dd \ln \Gamma}{\dd x} + g \chi_\rho \frac{P}{\rho} \frac{\dd \ln \chi_\rho}{\dd x} + g \chi_\rho \frac{P}{\rho} \frac{\dd \ln P}{\dd x}\\
        & \quad - g \chi_\rho \frac{P}{\rho} \frac{\dd \ln \rho}{\dd x} = [g_\Gamma \Gamma_\rho + ( -1 + \chi_\rho + \chi_{\rho \rho}) g ] \chi_\rho \frac{P}{\rho} \frac{\dd \ln \rho}{\dd x} \\
        & \quad + [g_\Gamma \Gamma_T + ( \chi_T + \chi_{\rho T}) g ] \chi_\rho \frac{P}{\rho} \frac{\dd \ln T}{\dd x}.
    \end{aligned}
\eeq
Thus, subtracting $(-1+\chi_\rho) g P/\rho$ times the mass equation~(\ref{eq:hydro_mat_mass_final}) from the energy equation~(\ref{eq:hydro_mat_en}) yields
\beq
    \begin{aligned}
        & (1 - \chi_\rho ) g \frac{P}{\rho} \frac{\dd \ln v}{\dd x} +  \left[ ( g_\Gamma \Gamma_\rho + g \psi_\rho ) \epsilon + ( g_\Gamma \Gamma_\rho - g ) \chi_\rho \frac{P}{\rho} \right] \frac{\dd \ln \rho }{\dd x} \\
        & \quad + \left[ ( g_\Gamma \Gamma_T + g \psi_T ) \epsilon + g_\Gamma \Gamma_T \chi_\rho \frac{P}{\rho} \right] \frac{\dd \ln T }{\dd x} = 0, 
    \end{aligned}
    \label{eq:hydro_mat_en_final}
\eeq
which determines the third row of matrix $\boldsymbol{\mathsf{M}}$ as
\beq
    \begin{aligned}
        m_{31} &= ( 1 - \chi_\rho ) g \frac{P}{\rho}, \quad m_{32} = (g_\Gamma \Gamma_\rho + g \psi_\rho ) \epsilon + ( g_\Gamma \Gamma_\rho - g ) \chi_\rho \frac{P}{\rho}, \\
        m_{33} &= ( g_\Gamma \Gamma_T + g \psi_T ) \epsilon + g_\Gamma \Gamma_T \chi_\rho \frac{P}{\rho}. 
    \end{aligned}
\eeq

\section{Derivation and solution of 1D hydrodynamic equations for ideal gas}\label{sec:hydro_id}

Using equations (\ref{eq:id}) the quantities defined by equations~(\ref{eq:Gamma_def}--\ref{eq:thermo_def}) are
\begin{equation}
    \Gamma = {\rm{const}}, \quad \! \! \chi_\rho = \chi_T = \psi_T = 1, \quad \! \! \Gamma_\rho = \Gamma_T = \chi_{\rho \rho} = \chi_{\rho T} = \psi_\rho = 0.
\end{equation}
We see that the equation~(\ref{eq:hydro_mat_mass_final}) reduces to equation~(\ref{eq:hydro_id_mass}). The momentum equation~(\ref{eq:hydro_mat_mom_final}) reduces to 
\begin{equation}
    \lp v^2 - g \frac{P}{\rho}\rp \frac{\dd \ln v}{\dd x} + g \frac{P}{\rho} \frac{\dd \ln T}{\dd x} = - \frac{\dd \phi_{\rm R}}{\dd x},
    \label{eq:hydro_id_mom_undone}
\end{equation}
and the energy equation (\ref{eq:hydro_mat_en_final}) to
\begin{equation}
    -g \frac{P}{\rho} \frac{\dd \ln \rho}{\dd x} + g \epsilon \frac{\dd \ln T}{\dd x} = 0.
    \label{eq:hydro_id_en_undone}
\end{equation}
By combining the expressions for $P$ and $\epsilon$ (equation \ref{eq:id}) with equation~(\ref{eq:hydro_id_en_undone}), we obtain equation~(\ref{eq:hydro_id_en}). Additionally, combining the expression for $g$ function (equation \ref{eq:g_def}) with equation~(\ref{eq:hydro_id_mom_undone}) and using already established equations gives the momentum equation (\ref{eq:hydro_id_mom}).

From the mass and energy equations~(\ref{eq:hydro_id_mass},\ref{eq:hydro_id_en}) it follows that
\begin{equation}
    v \rho T = v_0 \rho_0 T_0 = {\rm{const}}, \quad \frac{T}{\rho^{\Gamma-1}} = \frac{T_0}{\rho_0^{\Gamma-1}} = {\rm{const}}.
    \label{eq:hydro_id_scaling_poly}
\end{equation}
These equations also imply
\begin{equation}
    v \rho^\Gamma = v_0 \rho_0^\Gamma = {\rm{const}}, \quad T v^{\frac{\Gamma-1}{\Gamma}} = T_0 v_0^{\frac{\Gamma-1}{\Gamma}} = {\rm{const}}.
    \label{eq:hydro_id_scaling}
\end{equation}
Hence, from the momentum equation~(\ref{eq:hydro_id_mom}) we can derive
\begin{equation}
    \frac{\dd \ln v}{\dd x} = - \frac{1}{v^2-c_T^2} \frac{\dd \phi_{\rm R}}{\dd x}.
    \label{eq:hydro_id_mom_sol}
\end{equation}
Consequently, the critical speed at L1 ($\dd \phi_{\rm R}/\dd x = 0$) is the isothermal sound speed $v^2 (x_1) = c_T^2 (x_1) = k T (x_1)/\overline{m}$. Using the second scaling relation in equation~(\ref{eq:hydro_id_scaling}) allows us to solve equation~(\ref{eq:hydro_id_mom_sol}) by integration
\begin{equation}
    \int_{x_0}^x \left[ v - \frac{1}{v} \lp \frac{v_0}{v}\rp^{\frac{\Gamma-1}{\Gamma}} c_0^2 \right] dv = - \int_{x_0}^x d \phi_{\rm R},
\end{equation}
which leads to the solution for $v$, $\rho$, and $T$ profiles in the case of ideal gas as described by the set of equations~(\ref{eq:id_sol}).

To arrive at $v_0$, we evaluate equation~(\ref{eq:id_v}) at $x_1$ and use the expression
\begin{equation}
    v(x_1) = c_0^{\frac{2\Gamma}{3\Gamma-1}} v_0^{\frac{\Gamma-1}{3\Gamma-1}},
\end{equation}
which follows from the second scaling relation in equation~(\ref{eq:hydro_id_scaling}) and the condition for critical velocity $v (x_1) = c_T (x_1)$. In this way, we recover equation~(\ref{eq:id_v_0}). From the second scaling relation in equation~(\ref{eq:hydro_id_scaling_poly}) and ideal gas assumption (equation \ref{eq:id}) it follows that $P \propto \rho^\Gamma$. This is not surprising, because we start with a set of Euler equations neglecting radiation and any energy sink/source terms. Therefore, the MT is adiabatic and can be described by a polytrope.

\section{Mass-transfer rate comparison in the case of ideal gas}\label{sec:id_Mdot_comp_app}
The MT rate in the optically-thick regime described by equation~(\ref{eq:Mdot_KR}) for a polytropic donor can be calculated as
\begin{equation}
    \begin{aligned}
        &\dot{M}_{\rm{KR}} = -\frac{2\pi}{\sqrt{BC}} \int_{\bar{P} (R_{\rm L})=\bar{P}_1}^{\bar{P}(R_{\rm d})=0} F_3  \lp\frac{\bar{P}}{\bar{\rho}}\rp^{\frac{1}{2}} \dd \bar{P} \\
        & \: = \frac{2\pi}{\sqrt{BC}} K^{\frac{3}{2}} \Gamma F_3  \int_0^{\bar{\rho}_1} \bar{\rho}^{\frac{3}{2}( \Gamma-1)} \dd \bar{\rho} = \frac{2\pi}{\sqrt{BC}} K^{\frac{3}{2}} \frac{2 \Gamma F_3 }{3\Gamma-1} \bar{\rho}_1^{\frac{3\Gamma-1}{2}},
    \end{aligned}
    \label{eq:id_Mdot_M}
\end{equation}
where $F_3$ is given by equation~(\ref{eq:F3}) and $\bar{\rho}_1$ by equation~(\ref{eq:id_rho_stat}). The MT rate given by equation~(\ref{eq:Mdot_id}) in our model in the case of (adiabatic) ideal gas can be computed as
\begin{equation}
    \dot{M}_{\rm{thick}} = \frac{2\pi}{\sqrt{BC}} c_0^3 \frac{v_0}{c_0} \rho_0 = \frac{2\pi}{\sqrt{BC}} K^{\frac{3}{2}} \rho_0^{\frac{3\Gamma-1}{2}} \frac{v_0}{c_0},
    \label{eq:id_Mdot_thick}
\end{equation}
where $c_0^2 = K \rho_0^{\Gamma-1}$. Using equation~(\ref{eq:id_rho}) and the condition for the critical velocity $v^2 (x_1) = K \rho_1^{\Gamma-1}$ gives
\begin{equation}
    \frac{v_0}{c_0} = \lp \frac{\rho_1}{\rho_0} \rp^{\frac{3\Gamma-1}{2}}.
    \label{eq:v_0_c_0_frac}
\end{equation}
Together with equation~(\ref{eq:id_v_0}), we have for the density $\rho_1$ at the $x_1$ point
\begin{equation}
    \frac{3\Gamma-1}{2 (\Gamma-1)} \lp \frac{\rho_1}{\rho_0}\rp^{\Gamma-1} - \frac{1}{2} \lp \frac{\rho_1}{\rho_0}\rp^{3\Gamma-1} = \kappa_{\rm{max}} - \kappa, 
    \label{eq:id_rho1}
\end{equation}
Finally, dividing equations~(\ref{eq:id_Mdot_thick}) and (\ref{eq:id_Mdot_M}) and using equation~(\ref{eq:v_0_c_0_frac}) gives equation~(\ref{eq:id_Mdot_comp}).

Now, we want to compute the limits $\dot{\mathcal{M}}_0$ and $\dot{\mathcal{M}}_{\rm{max}}$ shown in equations~(\ref{eq:id_Mdot_comp_lim_0}) and (\ref{eq:id_Mdot_comp_lim_max}). For both hydrostatic density $\bar{\rho}_1$ and hydrodynamic density $\rho_1$  we  have $\bar{\rho}_1 \rightarrow \rho_0$ and $\rho_1 \rightarrow \rho_0$ as $\kappa \rightarrow 0^+$. Therefore, the limit $\dot{\mathcal{M}}_0$ is trivial. For the limit $\dot{\mathcal{M}}_{\rm{max}}$, the calculation is somewhat more complicated. Using equations~(\ref{eq:id_rho_stat}) and (\ref{eq:id_rho1}) we can derive
\begin{equation}
    \lp \frac{\rho_1}{\bar{\rho}_1}\rp^{\Gamma-1} \left[\frac{3\Gamma-1}{2(\Gamma-1)} - \frac{1}{2} \lp \frac{\rho_1}{\rho_0}\rp^{2\Gamma}\right] = \kappa_{\rm{max}} = \frac{\Gamma}{\Gamma-1}.
\end{equation}
Using the fact that $\rho_1 / \rho_0 \rightarrow 0$ as $\kappa \rightarrow \kappa_{\rm{max}}^-$ we arrive at
\begin{equation}
    \lim_{\kappa \to \kappa_{\rm{max}}^-} \frac{\rho_1}{\bar{\rho}_1} = \lp \frac{2\Gamma}{3\Gamma-1}\rp^{\frac{1}{\Gamma-1}}.
\end{equation}
Combining this result with the expression for the MT ratio in equation~(\ref{eq:id_Mdot_comp}) and using the definition of $F_3$ (equation \ref{eq:F3}), we get equation~(\ref{eq:id_Mdot_comp_lim_max}). Further, using the limit,
\begin{equation}
    \lim_{x \to 0} (1+r_1x)^{\frac{r_2}{x}} = e^{r_1 r_2},
    \label{eq:lim_e}
\end{equation}
we can compute the limit of $F_3$  as 
\begin{equation}
    \lim_{\Gamma \to 1} F_3 (\Gamma) = \lim_{\Gamma \to 1} \Gamma^{\frac{1}{2}} \lp 1 - \frac{\Gamma-1}{\Gamma+1}\rp^{\frac{\Gamma+1}{2(\Gamma-1)}} = \frac{1}{\sqrt{e}}.
    \label{eq:F3_lim}
\end{equation}
Using this result we can see that $\lim_{\Gamma \to 1} \dot{\mathcal{M}}_0 = \sqrt{e}$. Similarly, we can calculate
\begin{equation}
    \lim_{\Gamma \to 1} \dot{\mathcal{M}}_{\rm{max}} = \lim_{\Gamma \to 1} \Gamma^{-\frac{1}{2}} \left[1+\frac{(\Gamma-1)^2}{3\Gamma-1}\right]^{\frac{\Gamma+1}{2(\Gamma-1)}} = e^0 = 1.
\end{equation}

To truly prove that equation~(\ref{eq:id_Mdot_comp}) reduces to equation~(\ref{eq:iso_Mdot_comp}) in the limit $\Gamma \rightarrow 1$, we evaluate
\begin{equation}
    \dot{\mathcal{M}}_1 \equiv \lim_{\Gamma \to 1} \frac{\dot{M}_{\rm{thick}}}{\dot{M}_{\rm{KR}}}.
    \label{eq:M1_def}
\end{equation}
Using equation (\ref{eq:id_rho_stat}) and equation~(\ref{eq:lim_e}) we obtain
\begin{equation}
    \lim_{\Gamma \to 1} \frac{\bar{\rho}_1}{\rho_0} = \lim_{\Gamma \to 1} \lp 1 - \frac{\Gamma-1}{\Gamma} \kappa\rp^{\frac{1}{\Gamma-1}} = \exp (-\kappa).
    \label{eq:id_rho_stat_lim}
\end{equation}
We can rearrange equation~(\ref{eq:id_v_0}) into the form
\begin{equation}
    \frac{\Gamma}{\Gamma-1} \left\{ \left[ \lp\frac{3\Gamma-1}{2\Gamma}\rp^{\frac{1}{\Gamma-1}} \lp \frac{v_0}{c_0}\rp^{\frac{2}{3\Gamma-1}} \right]^{\Gamma-1} - 1\right\} = \frac{1}{2} \lp \frac{v_0}{c_0} \rp^2 - \kappa.
    \label{eq:id_v_0_rearr}
\end{equation}
Using equation~(\ref{eq:lim_e}) we calculate
\begin{equation}
    \lim_{\Gamma \to 1} \lp\frac{3\Gamma-1}{2\Gamma}\rp^{\frac{1}{\Gamma-1}} = \lim_{\Gamma \to 1} \lp 1 + \frac{\Gamma-1}{2\Gamma}\rp^{\frac{1}{\Gamma-1}} = \sqrt{e}.
\end{equation}
Therefore, using the identity $\lim_{x \to 0} (a^x-1)/x = \ln a$, for $a>0$, the left-hand side of the equation~(\ref{eq:id_v_0_rearr}) becomes $\ln (\sqrt{e} v_0/c_0)$ and equation~(\ref{eq:id_v_0_rearr}) reduces to
\begin{equation}
    \frac{1}{2} + \ln \frac{v_0}{c_0} = \frac{1}{2} \lp \frac{v_0}{c_0}\rp^2 -\kappa,
    \label{eq:id_v_0_lim}
\end{equation}
which identical to equation~(\ref{eq:hydro_iso_v_0}) provided that $c_0 = c_T$. Therefore, combining equations~(\ref{eq:F3_lim}), (\ref{eq:id_rho_stat_lim}), and (\ref{eq:id_Mdot_comp}) we get
\begin{equation}
    \dot{\mathcal{M}}_1 = \sqrt{e} \exp (\kappa) \frac{v_0}{c_0} = \exp \left[ \frac{1}{2} \lp \frac{v_0}{c_0}\rp^2 \right],
\end{equation}
If $c_0 = c_T$, equation~(\ref{eq:M1_def}) reduces to the MT ratio in the case of isothermal gas in equation~(\ref{eq:iso_Mdot_comp}).


\bsp	
\label{lastpage}
\end{document}